\documentclass[longauth]{aa}   

\usepackage{graphicx}
\usepackage{float}
\usepackage{natbib}
\usepackage{txfonts}
\usepackage{hyperref}
\usepackage{xspace}
\hypersetup{
    colorlinks=true,
    linkcolor=blue,
    filecolor=blue,      
    urlcolor=blue,
    citecolor=blue}
\usepackage{capt-of}
\usepackage{caption}
\usepackage[switch]{lineno}
\usepackage[svgnames]{xcolor}
\usepackage{orcidlink}

\linenumbers

\newcommand{\cii}{[C\,{\sc ii}]\xspace}

\newcommand{\unit}[1]{\mathrm{#1}}

\newcommand{\km}{\unit{km}}
\newcommand{\s}{\unit{s}}
\newcommand{\kms}{\km\,\s^{-1}}

\begin{document}

   \title{The ALPINE-CRISTAL-JWST survey}

   \subtitle{The Tully-Fisher Relation at $z = 4-6$.}

   \author{D. A. Gómez-Espinoza\orcidlink{0009-0008-1835-7557}\inst{1,2}
   \thanks{diego.gomeze@postgrado.uv.cl},
           J. Molina\orcidlink{0000-0002-8136-8127}\inst{1,2},
           E. Ibar\orcidlink{0009-0008-9801-2224}\inst{1,2},
          M. Aravena\orcidlink{0000-0002-6290-3198}\inst{3,2},
          R. J. Assef\orcidlink{0000-0002-9508-3667}\inst{3},
          M. Boquien\orcidlink{0000-0003-0946-6176}\inst{4},
          M. B\'ethermin\orcidlink{0000-0002-3915-2015}\inst{5},
          P. Cassata\orcidlink{0000-0002-6716-4400}\inst{6,7},
          E. da Cunha\orcidlink{0000-0001-9759-4797}\inst{8},
          P. Dam\orcidlink{0009-0007-7842-9930}\inst{6},
          M. Dessauges-Zavadsky\orcidlink{0000-0003-0348-2917}\inst{9},
          A. L. Faisst\orcidlink{0000-0002-9382-9832}\inst{10},
          S. Fujimoto\orcidlink{0000-0001-7201-5066}\inst{11,12},
          J. González-López\inst{13,14},
          R. Herrera-Camus\orcidlink{0000-0002-2775-0595}\inst{15,2},
          H. Inami\orcidlink{0000-0003-4268-0393}\inst{16},
          A. M. Koekemoer\orcidlink{0000-0002-6610-2048}\inst{17},
          L. Lee\orcidlink{0000-0001-7457-4371}\inst{18},
          Y. Li\orcidlink{0000-0001-8237-1515}\inst{19},
          L. Liu\orcidlink{0009-0004-1270-2373}\inst{10},
          B. Mobasher\orcidlink{0000-0001-5846-4404}\inst{20},
          M. Palla\orcidlink{0000-0002-3574-9578}\inst{21},
          M. Relano\orcidlink{0000-0003-1682-1148}\inst{22,23},
          M. Romano\orcidlink{0000-0002-9948-3916}\inst{24,7},
          J. D. Silverman\orcidlink{0000-0002-0000-6977}\inst{25,26,27,28},
          J. Spilker\orcidlink{0000-0003-3256-5615}\inst{19},
          H. Übler\orcidlink{0000-0003-4891-0794}\inst{18},
          D. Vergani\orcidlink{0000-0003-0898-2216}\inst{29},
          V. Villanueva\orcidlink{0000-0002-5877-379X}\inst{3,2},
          W. Wang\orcidlink{0000-0002-7964-6749}\inst{10} and
          G. Zamorani\orcidlink{0000-0002-2318-301X}\inst{29}
          }

   \institute{Instituto de F\'{i}sica y Astronom\'{i}a, Universidad de Valpara\'{i}so, Avda. Gran Breta\~{n}a 1111, Valpara\'{i}so, Chile.
   \and {Millennium Nucleus for Galaxies (MINGAL).} 
   \and {Instituto de Estudios Astrof\'{\i}sicos, Facultad de Ingenier\'{\i}a y Ciencias, Universidad Diego Portales, Av. Ej\'ercito 441, Santiago 8370191, Chile.}
   \and {Université Côte d'Azur, Observatoire de la Côte d'Azur, CNRS, Laboratoire Lagrange, 06000, Nice, France.}
   \and {Universit\'e de Strasbourg, CNRS, Observatoire astronomique de Strasbourg, UMR 7550, 67000 Strasbourg, France.}
   \and {Dipartimento di Fisica e Astronomia Galileo Galilei Universit{\`a} degli Studi di Padova, Vicolo dell’Osservatorio 3, 35122 Padova Italy.}
   \and {INAF - Osservatorio Astronomico di Padova, Vicolo dell’Osservatorio 5, I-35122, Padova, Italy.}
   \and {International Centre for Radio Astronomy Research (ICRAR), The University of Western Australia, M468, 35 Stirling Highway, Crawley, WA 6009, Australia.}
   \and {Département d’Astronomie, Université de Genéve, Chemin Pegasi 51, 1290 Versoix, Switzerland.}
   \and {IPAC, California Institute of Technology, 1200 E. California Blvd. Pasadena, CA 91125, USA.}
   \and {Department of Astronomy, The University of Texas at Austin, Austin, TX 78712, USA.}
   \and {David A. Dunlap Department of Astronomy and Astrophysics, University of Toronto, 50 St. George Street, Toronto, Ontario, M5S 3H4, Canada.}
   \and {Instituto de Astrof\'isica, Facultad de Física, Pontificia Universidad Cat\'{o}lica de Chile, Santiago 7820436, Chile 3 Las Campanas.}
   \and {Observatory, Carnegie Institution of Washington, Ra\'{u}l Bitr\'{a}n 1200, La Serena, Chile.}
   \and {Departamento de Astronomía, Universidad de Concepción, Barrio Universitario, Concepción, Chile.}
   \and {Hiroshima Astrophysical Science Center, Hiroshima University, 1-3-1 Kagamiyama, Higashi-Hiroshima, Hiroshima 739-8526, Japan.}
   \and {Space Telescope Science Institute, 3700 San Martin Drive, Baltimore, MD 21218, USA.}
   \and {Max-Planck-Institute für extraterrestrische Physik, Giessenbachstrasse 1, 85748 Garching, Germany.}
   \and {Department of Physics and Astronomy and George P. and Cynthia Woods Mitchell Institute for Fundamental Physics and Astronomy, Texas A$\&$M University, 4242.}
   \and {Department of Physics and Astronomy, University of California, Riverside, 900 University Ave, Riverside, CA 92521, USA.}
   \and{INAF – Osservatorio Astrofisico di Arcetri, Largo E. Fermi 5, 50125 Firenze, Italy.}
   \and {Dept. Física Te\'{o}rica y del Cosmos, Campus de Fuentenueva, Edificio Mecenas, Universidad de Granada, E-18071, Granada, Spain.}
   \and {Instituto Universitario Carlos I de Física Te\'{o}rica y Computacional, Universidad de Granada, 18071, Granada, Spain.}
   \and {Max-Planck-Institut für Radioastronomie, Auf dem Hügel 69, 53121 Bonn, Germany.}
   \and {Kavli Institute for the Physics and Mathematics of the Universe (Kavli IPMU, WPI), UTIAS, Tokyo Institutes for Advanced Study, University of Tokyo, Chiba, 277-8583, Japan.}
   \and {Department of Astronomy, Graduate School of Science, The University of Tokyo, 7-3-1 Hongo, Bunkyo, Tokyo 113-0033, Japan.}
   \and {Center for Data-Driven Discovery, Kavli IPMU (WPI), UTIAS, The University of Tokyo, Kashiwa, Chiba 277-8583, Japan.}
   \and {Center for Astrophysical Sciences, Department of Physics \& Astronomy, Johns Hopkins University, Baltimore, MD 21218, USA.}
    \and {INAF – Osservatorio di Astrofisica e Scienza dello Spazio di Bologna, Via Gobetti 93/3, 40129 Bologna, Italy.}
   }

   \date{Received  XXXXX; accepted YYYYY}

 
  \abstract{The Tully-Fisher relation provides a fundamental link between galaxy mass and their kinematics. Although its evolution is relatively well constrained up to cosmic noon, observational probes at higher redshifts remain scarce. In this work, we investigate the stellar and baryonic Tully-Fisher relations (TFRs) at $z=4-6$ by analyzing JWST/NIRSpec IFU observations of 18 main-sequence star-forming galaxies belonging to the ALPINE-CRISTAL-JWST survey. { We apply a morpho-kinematic disk classification scheme, finding five disk-like galaxies and} a survey disk fraction of $0.28\pm0.11$. These disk galaxies span stellar masses of $10^{9.5-10.5}\,M_\odot$, { intrinsic rotational velocities of $91-127\,{\rm km\,s^{-1}}$}, and intrinsic velocity dispersions of { $66-129\,{\rm km\,s^{-1}}$}. For both versions of the TFR, we find that the ALPINE-CRISTAL-JWST disks lie below the local relations and are likely offset from those reported at cosmic noon ($z\approx2.3$). By fixing the TFR slope to the $z=0$ value, we measure zero-point offsets for the stellar-mass TFR of { $-0.96\pm0.18$ dex} and { $-0.50\pm0.18$ dex} relative to the local Universe and cosmic noon, respectively. For the baryonic TFR, we find zero-point offsets of { $-0.57\pm0.20$ dex} and { $-0.30\pm0.20$ dex} with respect to the local Universe and cosmic noon relations. Our measurements are consistent with the redshift evolution of the zero-points of both TFRs being driven by the evolution of dark matter halo properties with cosmic time, modulated by a reduced contribution of dark matter to the dynamical mass on the galaxy scale.}


\authorrunning{Gómez-Espinoza, Diego~A., et al.}
\titlerunning{The Tully-Fisher relation at $z=4-6$}
   \maketitle
   \nolinenumbers
   
%

\section{Introduction}

Understanding how galaxies assembled their mass and structure in the early Universe is a key goal of extragalactic astrophysics. During the first Gyr of cosmic history, typical star-forming galaxies were rapidly growing through gas accretion and mergers, while their internal structure was still settling \citep{Madau&Dickinson2014,Dayal&Ferrara2018,Lefevre+20,Romano+21,Herrera-Camus+25}. This early phase is key for studying how baryons assembled within dark matter (DM) haloes, and when { rotationally} supported galaxies began to emerge.

Gas kinematics provides a straightforward way to address the dynamical state of galaxies. { For example, integral field unit (IFU) surveys observing the ionized-gas kinematics of galaxies at $z\approx1-3$} revealed a diverse galaxy population, including turbulent rotating disks, dispersion-dominated systems, and mergers \citep{forster-schreiber+09,Wisnioski+15}. { These galaxies were found, on average, exhibiting larger velocity dispersions and lower rotational support compared to their local counterparts} \citep{Wisnioski+15,Genzel+17,Price+20}. { However, the magnitude of this evolution remains debated and appears to depend on the gas tracer. Measurements based on cold gas generally show lower velocity dispersions and higher rotational support than ionized-gas measurements, implying a milder evolution of disk turbulence and rotational support with redshift} \citep{Girard+21,Rizzo+21,Roman-Oliveira+23,Rizzo+24}. Extending such analysis to earlier epochs is crucial to determine when stable rotating disks become common, how disk settling proceeds with cosmic time, and how this process is connected to the build-up of stellar mass ($M_\star$) within galaxies.

In this context, a useful way to study the connection between the mass of a galaxy and its dynamics is the Tully-Fisher relation (TFR). Originally defined as a relationship between the luminosity and rotation velocity of systems \citep{Tully&Fisher77}, it is now commonly studied in terms of their stellar mass and/or baryonic mass ($M_{\rm bar}$) versus a characteristic circular velocity (e.g., \citealt{Mcgaugh+00,Reyes+11,Swinbank+12,Lelli+16}). The stellar-mass TFR (sTFR) and the baryonic TFR (bTFR) provide a powerful framework for testing how disk-like galaxies assemble their mass within dark matter halos. The cosmic evolution of both the sTFR and bTFR can place constraints on gas fractions, star-formation efficiency, feedback, and the relative contributions of baryons and dark matter on galaxy scales (e.g., \citealt{Ubler+17,tiley+19,Birkin+24,Sharma+24}).

Observationally, the cosmic evolution of the TFR has been studied extensively up to $z\approx2$, with mixed results reported in the literature depending on the redshift range, sample selection, and galaxy population. Some studies report little or no evolution in the sTFR out to $z\approx1-2$, suggesting that the sTFR was already in place for massive star-forming disks at relatively early times \citep{Conselice+05,Miller+11,DiTeodoro+16,tiley+19,Birkin+24}. In contrast, other studies report negative offsets with respect to the local relation, indicating that galaxies at fixed stellar mass show higher circular velocities at earlier cosmic times, possibly due to an evolution in the efficiency of gas-to-star conversion within growing dark matter halos \citep{Puech+08,Vergani+12,Ubler+17,Harrison+17}. { Overall, there is currently no clear consensus on the cosmic evolution of the sTFR. The disagreement among the studies can arise from observational and methodological differences in the derivation of galaxy kinematics, stellar masses, and data quality} \citep{Ubler+17,Pelliccia+19,Sharma+24,Espejo-Salcedo+25,Mancera-Pina+26}. Compared to the sTFR, the bTFR cosmic evolution remains less constrained { because robust measurements of gas masses are required, which remain challenging at high redshift}. In the local Universe, adding the gas mass into the mass budget for dwarf galaxies moves these systems into a single baryonic mass TFR \citep{Mcgaugh+00,Lelli+16}. At high redshift, accounting for gas mass becomes increasingly important for massive systems { because galaxy gas fractions increase toward earlier cosmic times} \citep{Tacconi+20}.

Probes of both the bTFR and the sTFR at $z \gtrsim 4$ remain scarce, mainly due to observational limitations. Recent Atacama Large Millimeter/sub-millimeter Array (ALMA) observations of the \cii\,158~$\mu$m line emission have revealed that some galaxies at $z\approx4-6$ are consistent with being rotating disks \citep{Lefevre+20,jones+21,Herrera-Camus+22,Lee+25}. However, observations remain limited because long exposure times are required to properly spatially resolve the systems. The James Webb Space Telescope (JWST) has opened a new window on this problem. Early Near Infrared Camera (NIRCam) imaging studies showed that galaxies at $z\approx4-6$ span a wide range of morphologies, including disturbed systems, mergers, and a significant fraction of disk-like objects \citep{Ferreira+22,Kartaltepe+23,Lee+24}. Some of the more massive systems even show bulges in the early stages of formation \citep{Huertas-Company+24}. JWST IFU observations now make it possible to map rest-frame optical emission lines, such as H$\alpha$, on kpc scales at these epochs, allowing for a direct comparison between ionized-gas kinematics, stellar morphology, and cold-gas tracers. In parallel, similar multi-wavelength approaches are being extended even deeper into the end of the epoch of reionization \citep{Algera+25,Fisher+26}. This new observational landscape makes it possible to explore the TFR at $z \gtrsim 4$, in an epoch where galaxies are expected to be more gas-rich, more turbulent, and still assembling a significant fraction of their stellar and baryonic mass. Studying the TFR in this regime is therefore important because it probes the epoch when the connection between galaxy mass and gravitational potential may still be emerging, offering direct constraints on the early build-up of disks and on the relative role of baryons and dark matter in young galaxies.

In this paper, we use JWST/Near Infrared Spectrograph (NIRSpec) IFU observations towards a sample of main-sequence star-forming galaxies at $z = 4-6$, the ALPINE-CRISTAL-JWST survey \citep{Faisst+26}. The IFU data enable the characterization of the ionized-gas kinematics of galaxies at high resolution ($0\farcs15 \sim 1$ kpc at $z = 5$), allowing more robust kinematic measurements than single-slit or slitless spectroscopy. Combined with JWST/NIRCam images, the JWST/NIRSpec IFU data allow us to identify disk-like systems. Our main goal is to study the Tully-Fisher relation in its baryonic and stellar-mass versions, and to explore possible cosmic evolution of the TFRs through comparisons with lower-$z$ samples (e.g., \citealt{Reyes+11,Lelli+16,Ubler+17}). 

The paper is organized as follows: Section~\ref{sec:data} describes the sample, observations, and ancillary data used. In Section~\ref{sec:methods}, we introduce the models adopted to analyze the data and our galaxy disk classification scheme. We present the baryonic and stellar-mass TFRs in Section~\ref{sec:results} and discuss them in Section~\ref{sec:discussion} by means of comparisons with other observational works and with a simple empirical model. Finally, we conclude in Section~\ref{sec:conclusions}. We adopt a standard flat $\Lambda$-CDM cosmology with $H_0 = 70\, {\rm km \ s}^{-1} {\rm Mpc}^{-1}$, $\Omega_{\rm M} = 0.3$, and $\Omega_{\Lambda} = 0.7$. 


\section{Observations and Data Reduction}
\label{sec:data}
We analyze JWST/NIRSpec IFU observations (Cycle-2 program GO--3045; P.I.: A. Faisst) of 18 main-sequence star-forming galaxies at $z = 4-6$, the ALPINE-CRISTAL-JWST survey. The targets were selected from the ALMA large programs ALPINE \citep{Lefevre+20,Bethermin+20,Faisst+20,Faisst+22} and CRISTAL \citep{Herrera-Camus+25}, which provide \cii observations at $0\farcs2-0\farcs7$ spatial resolution ($\approx 1-5$ kpc at $z = 5$). The ALPINE-CRISTAL-JWST survey benefits from the exquisite multi-wavelength imaging coverage of the COSMOS field (\citealt{Scoville+07,Weaver+22,Donnan+24,Shuntov+25}), which enables co-spatial comparisons with the stellar continuum observed at sub-kpc scales. This makes the ALPINE-CRISTAL-JWST sample a key reference to connect galaxy properties between the post-reionization era, cosmic noon, and the local universe. The sample spans stellar masses of $M_\star \approx 10^{9-11}\,{\rm M}_\odot$ and SFR $\approx 10-200\,{\rm M}_\odot$\,yr$^{-1}$, { as derived from spatially integrated SED fitting with CIGALE, combining ground-based and Spitzer photometry from COSMOS2020 with JWST/NIRCam imaging from COSMOS-Web. We refer to \citet{Mitsuhashi+24}, \citet{Herrera-Camus+25}, and \citet{Faisst+26} for full details of the SED-fitting procedure}. { High-resolution \cii kinematic analyses from \cite{Lee+25} are available for 16 out of 18 targets. The exceptions are DC--742174 (C--17) and VC--5110377875.} A summary of the global properties of the galaxies is given in Appendix~\ref{app:globalprop}.

\subsection{JWST/NIRSpec IFU}

The NIRSpec/IFU observation strategy and exposure times are outlined in \cite{Faisst+26}, while the data reduction pipeline used to derive science-quality data products is presented in \citet{Fujimoto+25}. The survey adopts the medium resolution disperser/filter combinations G235M/F170LP ($\sim1.7-3.2\,\mu$m) and G395M/F290LP ($\sim 2.9-5.3\,\mu$m), covering the main rest-frame optical emission lines (H$\alpha$, H$\beta$, [N\,{\sc ii}], [O\,{\sc ii}], [O\,{\sc iii}], and [S\,{\sc ii}]) at a spectral resolution of $R \approx 1000$ { ($\sim127$\,km\,s$^{-1}$ in velocity dispersion)}. The data reduction pipeline largely follows the standard JWST/STScI pipeline (v1.16.0), as described in \citet{Rigby+25}, plus the application of several post-processing steps to improve the final data quality. Those routines consider additional cleaning of residual detector artifacts, improved background subtraction, masking of intermittent light leakages, removal of residual striping patterns in collapsed continuum maps, estimation of the data uncertainties, and astrometry alignment of the data cubes to the COSMOS-Web NIRCam mosaics, which are tied to the Gaia reference frame. We refer to \citet{Faisst+26} and \citet{Fujimoto+25} for more details.

\subsubsection{DC--519281 / C--09}

In the specific case of DC--519281 (or C--09), we analyze NIRSpec/IFU observations obtained under the Origin of the [C\,{\sc ii}] Halos in Distant Systems (ORCHIDS) program (GO--5974; P.I.: M. Aravena). The main advantage of using ORCHIDS data is the improved spectral resolution due to the use of the G395H/F290LP disperser/filter combination ($R\approx2700$; $\sigma_{\rm LSF}\sim50$\,km\,s$^{-1}$), plus the two-times finer pixel scale produced by the dithering strategy applied during the observations. The ORCHIDS data reduction follows the same pipeline used to reduce the ALPINE-CRISTAL-JWST sample data \citep{Fujimoto+25}, as briefly outlined above. We provide a comparison for both datasets in Appendix~\ref{app:1}.

\subsection{JWST/NIRCam imaging}
All targets are within the COSMOS field, benefiting from exquisite multi-wavelength coverage. High-quality JWST/NIRCam images (F115W, F150W, F277W, and F444W filters) from COSMOS-Web (PID 1727, PI: J. Kartaltepe, \citealt{Casey+23}) are available for most of our targets (17/18). A subset of the sample is also covered by PRIMER (PID 1837, PI: J. Dunlop, \citealt{Donnan+24}), providing additional imaging coverage (F090W, F200W, F356W, and F410M filters). At $z = 4-6$, these JWST images probe the rest-frame UV/optical continuum. We use F444W image mosaics drizzled onto a common grid with a pixel scale of 0\farcs03. The F444W images trace the underlying stellar continuum (rest-frame $I-$band) of our targets. Full details on the JWST/NIRCam data reduction for the ALPINE-CRISTAL-JWST sample are given in \cite{Faisst+26}.

\section{Methods}
\label{sec:methods}
Our main goal is to study the Tully-Fisher relation, so we focus on carefully estimating the circular velocities ($V_{\rm circ}$) while adopting the stellar masses provided by \cite{Faisst+26} for our targets { (see Section~\ref{sec:data})}. We fit the data cubes pixel-by-pixel to build maps of H$\alpha$ emission-line intensity, line-of-sight velocity ($V_{\rm LOS}$), and velocity dispersion ($\sigma_V$). The kinematic data are then complemented with JWST F444W images to identify disk-like systems. For disk galaxies, we compute asymmetric drift correction terms and estimate $V_{\rm circ}$. Finally, to assess the effect of observational systematics we adopt a forward modeling approach.

\subsection{Imaging analysis}
\label{subsec:imaging_mod}
We model the JWST/NIRCam F444W images to characterize the stellar morphology of our targets, as those roughly correspond to rest-frame $I-$band data for galaxies at $z = 4-6$ \citep{Roper+22}. We adopt a single-component S\'ersic profile \citep{sersic63} and use {\sc pysersic} \citep{Pasha&miller23}. The image model has seven free parameters: $x/y$ centroids, photometric position angle (PA$_{\rm phot}$), effective radius ($R_e$), axis ratio ($q=b/a$), total flux, and S\'ersic index ($n$). The fitting routine considers the convolution by the JWST/NIRCam point-spread function (PSF) taken from the JWST's PSF simulation tool\footnote{\url{https://www.stsci.edu/jwst/science-planning/proposal-planning-toolbox/psf-simulation-tool}}. We note that this choice does not significantly affect the inferred parameters, as previous JWST/NIRCam studies have shown that simulated and empirical PSFs lead to consistent results \citep{Lee+24}. Prior to modeling, we mask spurious features (e.g., foreground stars and corner artifacts) manually. When fitting, we found that the S\'ersic indices were not well constrained due to the compactness of the sources and/or low S/N, but the best-fit values were typically close to unity { (average $n =1.03$, median of 0.95, and scatter of 0.47).} Therefore, we fixed $n=1$ (exponential profile). { Fixing $n=1$ produces only small changes in the morphological parameters, with median absolute differences of $7.2\%$, $1.6\%$, and $0.9\%$ in $R_e$, inclination, and PA$_{\rm phot}$, respectively. We show the quality of our S\'ersic models in Appendix~\ref{app:exponential_models}.\footnote{For DC-417567 we do not have available F444W imaging, so we used a pseudo F444W image from the datacube.}}

We assume a galactic oblate geometry to compute the inclination angles ($i$):

\begin{equation}
    \cos^2(i) = \frac{q^2-q_0^2}{1-q_0^2},
\end{equation}

\noindent where $q$ is the projected axis ratio ($b/a$) of the galaxy on the sky, and $q_0$ is the intrinsic axis ratio of the galaxy. We assume $q_0 = 0.2$, a value appropriate for describing moderately thick disks (see e.g., \citealt{Danhaive+25}). This assumption is further motivated by observational evidence that high-$z$ star-forming galaxies have thick disks \citep{Vanderwel+14}, as well as by cosmological simulations (e.g., \citealt{Pillepich+19}). Nevertheless, inferred kinematic quantities depend weakly on the exact choice of $q_0$ within commonly adopted values. For example, if the true intrinsic thickness were higher ($q_0 \approx 0.25-0.5$), the inclination-corrected velocities would only vary by $10\%$ \citep{Danhaive+25}, { keeping the main results of this work unchanged (Section~\ref{sec:results}).}

\begin{figure*}
    \centering
    \includegraphics[width=1\linewidth]{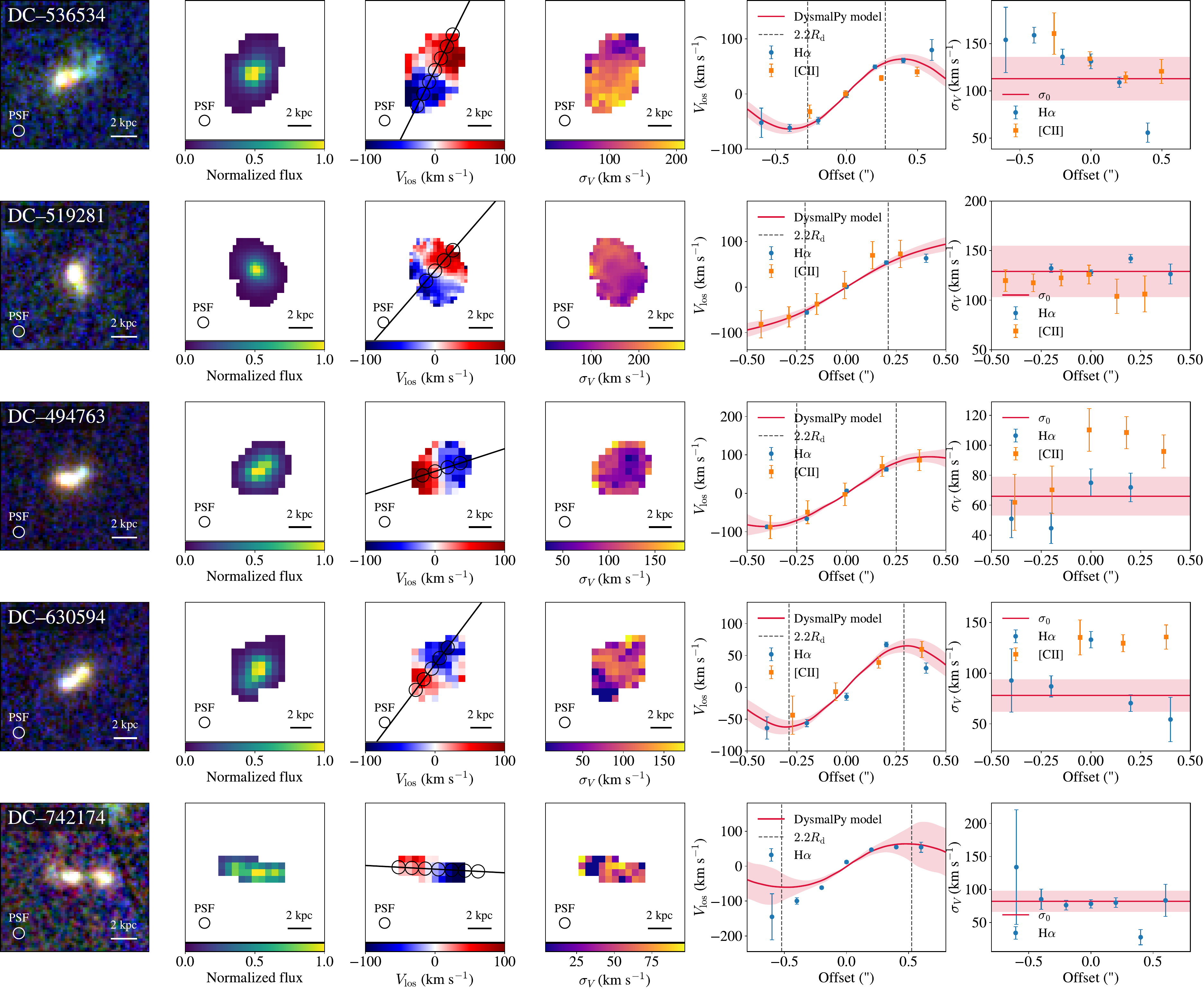}
    \caption{False-color JWST-NIRCam post-stamp image (Red: F444W, Green: F277W, Blue: F115W), NIRSpec-IFU H$\alpha$ intensity map, line-of-sight velocity field, velocity dispersion map, rotation curve, and velocity dispersion profile (\textit{columns}) for the five disk-like galaxies (\textit{rows}) in the ALPINE-CRISTAL-JWST sample. In the third column, the line indicates the kinematic position angle, and the grey circles mark apertures used to derive the rotation curve data. In the 5th column, the vertical dashed lines indicate the radius ($2.2\,R_{\rm d}$) at which the rotation velocity is extracted. { The red curve shows the best-fitting {\sc DYSMALpy} model after projection and convolution with the PSF and LSF, while the red-shaded region indicates its $1\sigma$ confidence interval. In both 5th and 6th columns, blue and orange points show the H$\alpha$ and [CII] measurements, respectively, and the error bars indicate $1\sigma$ uncertainties.}}
\label{fig:summary_5x5}

    \label{fig:placeholder}
\end{figure*}

\subsection{Emission line fitting}
\label{subsec:line_fitting}
We adopt the H$\alpha$ emission line as our main tracer of the ionized-gas kinematics, as it enables a fair comparison between our measurements and pre-JWST TFR studies. For each datacube, we fit pixel-by-pixel the H$\alpha$, [N,{\sc ii}],$\lambda \lambda$6548, 6584, [S,{\sc ii}],$\lambda \lambda$6717, 6731 emission lines with Gaussian models using the {\sc lmfit} package \citep{Newile+25} in {\sc Python}. The underlying stellar continuum is modeled with a linear function as the observations' S/N is found to be insufficient to detect any stellar absorption feature. The emission lines are fitted with single Gaussian components, tying their velocity centroid and velocity dispersion while leaving the individual line intensities free. { We assume all lines trace the same bulk ionized-gas kinematics (e.g., \citealt{Ho+16}). Hence, possible line-dependent centroid offsets or differences in linewidth are not captured by our model.} Observed line widths are corrected for instrumental line broadening by { considering} the NIRSpec line-spread function (LSF) { in convolution within the model}. We adopt the LSF width measurements of \cite{Jakobsen+22} for the grating/filter G395M/F290LP {($\sim 127$\,km\,s$^{-1}$ on average)}.

To select only the spectra that are well-characterized by the emission line models, we compare the best-fit solutions with a simple straight-line (featureless) model and compute the Bayesian Information Criterion difference ($\Delta$BIC, \citealt{Liddle+07}). We keep pixels with $\Delta$BIC$\,\equiv\,$BIC$_{\rm featureless} - $BIC$_{\rm Gaussian} > 10$ { corresponding to ``decisive evidence'' in favor of the Gaussian emission-line model over the featureless model} \citep{Liddle+07}. After this step, we remove isolated spurious pixel detections by applying the {\sc binary\_erosion} task of {\sc scipy.ndimage}\footnote{\url{https://docs.scipy.org/doc/scipy/reference/ndimage.html}} with a $3\times3$ pixel-box structure. However, as this task also erases good quality pixels characterizing high S/N emission from the galaxy boundaries, we recover those by applying the {\sc binary\_dilation} task of {\sc scipy.ndimage} { with the same pixel-box structure}. The best-fit parameters are then used to build the H$\alpha$ emission-line intensity, $V_{\rm LOS}$, and $\sigma_V$ two-dimensional maps (Fig.~\ref{fig:summary_5x5}). { Only spaxels satisfying $\Delta$BIC$>10$ and retained by the morphological mask described above are included in the two-dimensional maps. No additional S/N threshold is applied.}

\subsection{Disk classification}
\label{subsec:disk_class}
Given the relatively small sample size and the complex morphologies expected for galaxies at $z = 4-6$, we adopt a two-step classification scheme to identify disk-like galaxies in our sample. First, we visually inspect the H$\alpha$ velocity fields and the JWST/NIRCam F444W images for all 18 galaxies, focusing on identifying close companions within the NIRSpec/IFU field-of-view (FoV; $5'' \approx 30-35$\,kpc at $z =  4-6$). We classify our targets into ``galaxy pairs'' and ``isolated galaxy'', depending on whether a companion is present (or not) within the IFU field-of-view. We find that all the galaxy pairs show systemic velocity ($V_{\rm sys}$) difference of $\lesssim 150$\,km\,s$^{-1}$. Second, we apply the morpho-kinematic disk classification scheme of \cite{Wisnioski+15} to the isolated galaxies. Briefly, the disk classification scheme of \cite{Wisnioski+15} corresponds to five criteria that a source must satisfy for being labeled as ``disk galaxy'', namely:
\begin{enumerate}
    \item The velocity map exhibits a continuous velocity gradient along a single axis.
    \item The ratio between the rotational velocity (amplitude of the velocity gradient) and the velocity dispersion must be greater than or similar to one.
    \item The velocity dispersion map must peak near ($\lesssim$1 PSF FWHM) the kinematic center.
    \item { For inclined galaxies ($q <0.6$),} the photometric and kinematic position angles must agree within $< 30^\circ$.
    \item The position of the kinematic center coincides, within one PSF, with the centroid of the continuum.
\end{enumerate}
Photometric quantities are measured from F444W images (Section~\ref{subsec:imaging_mod}), while kinematic position angles (PA$_{\rm kin}$) are derived from H$\alpha$ $V_{\rm LOS}$ maps (Section~\ref{subsec:kin_pars}). We support the assessment of criterion 3 by considering available \cii velocity dispersion profiles from \citet{Lee+25}. The \cii line traces the cold, neutral gas component (see e.g. \citealt{Algera+25}); hence it provides an independent measurement of the velocity dispersion profile for the dominant mass component within our systems. We show the \cii velocity dispersion profiles as orange points in Figure~\ref{fig:summary_5x5}, { These profiles are extracted following the {DYSMALpy}-based procedure along PA$_{\rm kin}$ (see \citealt{Lee+25} for more details).}

\begin{figure}
    \centering
    \includegraphics[width=1\linewidth]{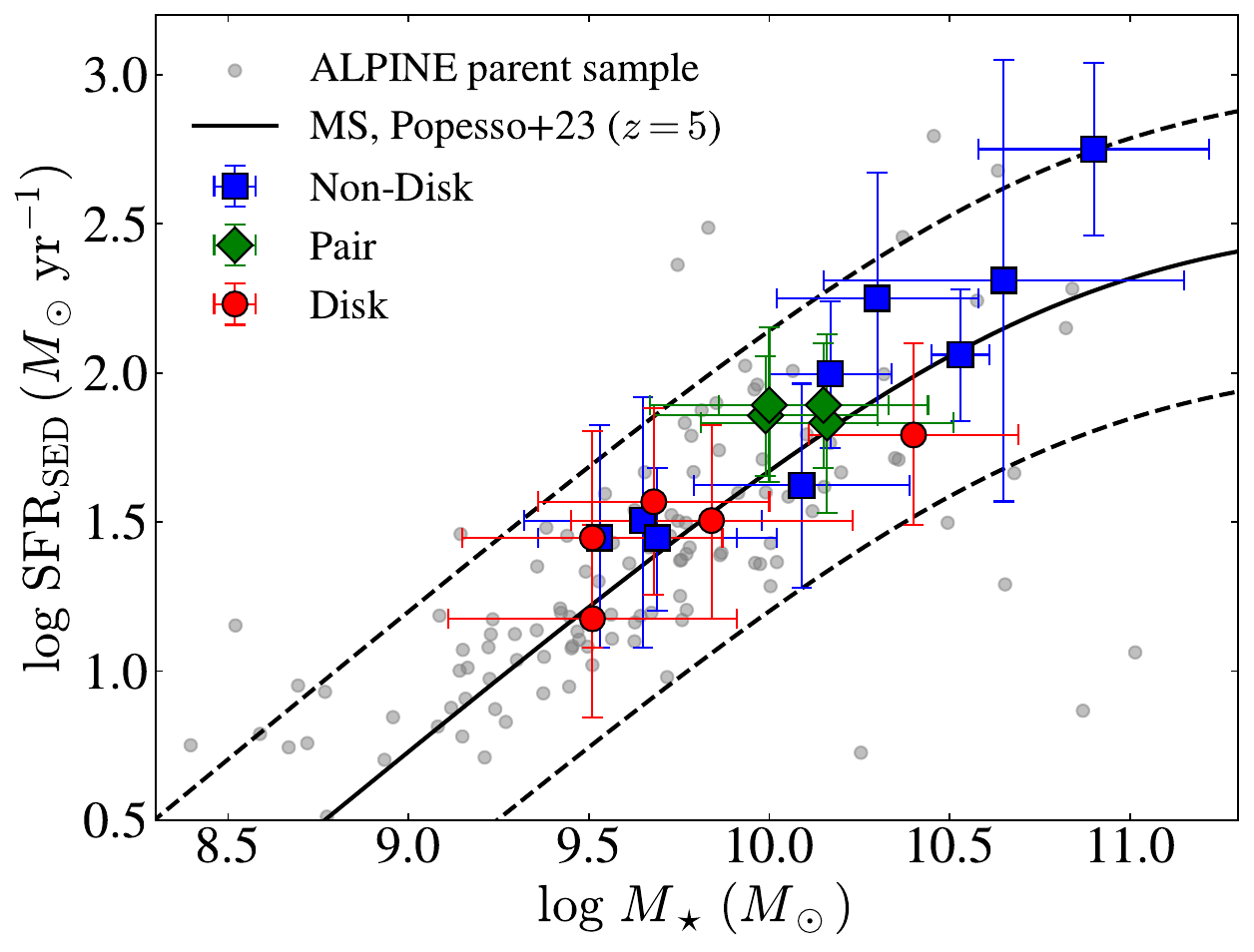}
    \caption{Star formation rate as a function of stellar mass for our sample. As a reference, we show the main-sequence of star-forming galaxies at $z = 5$ of \cite{Popesso+23}, with  the dashed lines indicating $3\times$ the main sequence}
    \label{fig:main-sequence}
\end{figure}

As a first step, we visually inspect the NIRSpec/IFU data cubes and the NIRCam/F444W images. We identify 14 isolated galaxies within the FoV of the IFU ($5'' \approx 30-35$\,kpc at $z =  4-6$). The other four systems correspond to galaxy pairs with $\Delta V_{\rm sys} \lesssim 150$\,km\,s$^{-1}$. Among the isolated galaxies, nine satisfy our first two disk-classification criteria, namely, they show a continuous velocity gradient aligned with a single axis and present a velocity amplitude-to-velocity dispersion ratio $\gtrsim 1$. The third criterion requires the velocity dispersion to peak near the region of steepest velocity gradient, which is expected to be close to the kinematic center for disk-like systems. Only for DC--630594 we observe that the velocity dispersion peaks at the kinematic center, while the rest of the galaxies show velocity dispersion radial profiles that can be roughly regarded as constant. In practice, the ionized-gas velocity dispersion can be affected by non-circular motions (e.g., outflows) and by local features \citep{Forster-schreiber+19,Phillips+25}, making the interpretation of the maps complex. Thus, to provide an additional assessment, we complement the analysis of the H$\alpha$ kinematics with an examination of the \cii velocity dispersion profiles presented in \cite{Lee+25}. The \cii line traces the total cold gas \citep{Dessauges-Zavadsky+20} and, although it could also be affected by outflows \citep{ginolfi+20,Romano+23,Birkin+25}, combining ionized gas and cold gas ($T \lesssim 100\,$K) information provides a more robust assessment of the dispersion peak. Using this approach, we find that seven galaxies satisfy the third criterion.
The fourth criterion requires, for galaxies with $q < 0.6$, consistency between the photometric and kinematic major axes (within $30^\circ$), which is fulfilled by six galaxies. Finally, we discard the system VC--5100994794 because it does not meet the fifth criterion. That is, its kinematic center does not coincide with the continuum peak seen in the F444W image. Therefore, we find that a total of five galaxies can be regarded as disks ($5/18 = 27.8\%$), nine systems show perturbed kinematics ($9/18 = 50.0\%$), and the remaining four targets ($4/18 = 22.2\%$) correspond to galaxy pairs. Figure~\ref{fig:main-sequence} shows our sample in the SFR--$M_\star$ plane separated by their morpho-kinematic classification (Table~\ref{tab:w15_criteria}). 

We note that two of the galaxies classified as disk-like in this work have been previously discussed as possible AGN and/or ionized-gas outflow hosts. In particular, DC--536534 was identified as a robust AGN candidate by \citet{Ren+25}. This target shows a broad H$\alpha$ emission line, regarded as a broad-line region component, plus a possible, additional outflow component. \citet{Ren+25} also classify DC--519281 as an AGN candidate, with possible evidence for an outflow component bright in [O\,{\sc iii}] emission. However, neither of these disk galaxies show significant cold-gas outflow signatures \citep{Birkin+25}.

We also note that DC--742174 is a borderline case in our disk classification scheme. { On one hand,} its rest-frame optical morphology shows two clump-like structures with no clear central emission. { On the other hand,} the H$\alpha$ velocity field displays a coherent velocity gradient along a single axis, the H$\alpha$ emission peaks in the galaxy center, and the observed velocity dispersion profile is broadly consistent with a roughly constant intrinsic dispersion. It is well-known that, at high-redshift, systems with irregular, often clumpy, morphologies in the rest-frame optical, can display regular velocity fields, reflecting an underlying disk nature (e.g., \citealt{Forster-schreiber&wuyts20}, and references therein). { However, given the limited spatial resolution of our observations, we cannot exclude the possibility that the observed kinematics arise from an unresolved interaction or from two systems projected close to each other in the sky. Therefore we flag DC--742174 separately throughout the analysis. Nevertheless, excluding DC--742174 from our analysis does not significantly change our results and conclusions (Section~\ref{sec:results}).}

\begin{table}
\centering
\def\arraystretch{1}
\setlength\tabcolsep{1.5pt}
\caption{Summary of the disk classification of the galaxies.}
\vspace{-0.2mm}
\label{tab:w15_criteria}
\begin{tabular}{lcccccccc}
\hline
ID & \multicolumn{5}{c}{Criteria} & \multicolumn{2}{c}{Type} \\
   & 1 & 2 & 3 & 4$^a$ & 5 & This work &  \cii$^b$ \\
\hline
DC--494763      & $\checkmark$ & $\checkmark$ & $\checkmark$ & $\checkmark$ & $\checkmark$ & Disk     & Best-Disk  \\
DC--519281      & $\checkmark$ & $\checkmark$ & $\checkmark$ & $\checkmark$ & $\checkmark$ & Disk     & Disk  \\
DC--536534      & $\checkmark$ & $\checkmark$ & $\checkmark$ & $\checkmark$ & $\checkmark$ & Disk     & Best Disk  \\
DC--630594      & $\checkmark$ & $\checkmark$ & $\checkmark$ & $\checkmark$ & $\checkmark$ & Disk     & Best Disk  \\
DC--742174$^{c}$      & $\checkmark$ & $\checkmark$ & $\checkmark$ & $\checkmark$ & $\checkmark$ & Disk     & ...  \\
DC--709575      & $\checkmark$ & $\checkmark$ & $\times$     & \ldots        & \ldots        & Non-Disk & Non-Disk  \\
DC--842313      & $\times$     & \ldots        & \ldots       & \ldots        & \ldots        & Non-Disk & Non-Disk  \\
DC--848185      & $\times$     & \ldots        & \ldots       & \ldots        & \ldots        & Non-Disk & Disk  \\
DC--873756      & $\times$     & \ldots        & \ldots       & \ldots        & \ldots        & Non-Disk & Non-Disk  \\
VC--5100541407  & $\checkmark$ & $\checkmark$ & $\times$     & \ldots        & \ldots        & Non-Disk & Non-Disk  \\
VC--5100994794  & $\checkmark$ & $\checkmark$ & $\checkmark$ & $\checkmark$  & $\times$      & Non-Disk & Non-Disk  \\
VC--5101218326  & $\times$     & \ldots        & \ldots       & \ldots        & \ldots        & Non-Disk & Non-Disk  \\
VC--5101244936  & $\times$     & \ldots        & \ldots       & \ldots        & \ldots        & Non-Disk & \ldots   \\
VC--5110377875  & $\checkmark$ & $\checkmark$ & $\checkmark$ & $\times$      & \ldots        & Non-Disk & \ldots  \\
DC--417567      & $\times$     & \ldots        & \ldots       & \ldots        & \ldots        & Pair     & \ldots  \\
DC--683613      & $\times$     & \ldots        & \ldots       & \ldots        & \ldots        & Pair     & Disk  \\
DC--873321      & $\times$     & \ldots        & \ldots       & \ldots        & \ldots        & Pair     & Disk  \\
VC--5100822662  & $\times$     & \ldots        & \ldots       & \ldots        & \ldots        & Pair     & Non-Disk  \\
\hline
\end{tabular}
{\flushleft { Notes.} { ($a$) Criterion 4 is applied only to galaxies with $q<0.6$, following \citet{Wisnioski+15}.} ($b$) Taken from \cite{Lee+25}. { ($c$) Its peculiar morphology could be indicative of a merger (Section~\ref{subsec:disk_class}).}}
\end{table}

\subsection{Kinematic measurements}
\label{subsec:kin_pars}

For each disk galaxy, we start by characterizing its kinematics by deriving the major kinematic axis of the $V_{\rm LOS}$ map. The major kinematic axis is set by PA$_{\rm kin}$, which we determine by using {\sc PAFIT} \citep{Krajnovic+06}. { PAFIT determines PA$_{\rm kin}$ by finding the axis that best reproduces the observed velocity field symmetry (see \citealt{Krajnovic+06} for more details)}. Here, we fix the kinematic center to the centroid of the H$\alpha$ emission, as these are broadly consistent with the F444W photometric centroid within a PSF-sized region (full-width at half-maximum $\sim 0\farcs15-0\farcs17$, \citealt{Deugenio+24}). 

To construct rotation curves and $\sigma_V$ radial profiles, we follow the method used in {\sc DYSMALpy} \citep{Price+21,Lee+2025}. Briefly, we use the kinematic center and PA$_{\rm kin}$ to derive the major kinematic axis. Across the major kinematic axis, we extract a set of integrated spectra using a series of circular apertures { with a diameter of $0\farcs2$ and separated by $0\farcs2$, avoiding overlapping apertures and reducing covariance between neighbouring measurements.} For each integrated spectrum, we fit the H$\alpha$ emission line to obtain its centroid and width, corresponding to the $V_{\rm LOS}$ and $\sigma_V$ values that we use to construct the rotation curve and the velocity dispersion radial profile (Fig.~\ref{fig:summary_5x5}). Here, we only consider best-fit results if the H$\alpha$ emission line has S/N$> 5$. 

From each rotation curve, we compute the inclination-corrected rotation velocity values by taking the average between $V_{\rm LOS}$ extracted at a radius $R = 2.2R_{\rm d}$, where $R_{\rm d}$ is the disk-scale length.\footnote{For an exponential disk we have $R_{\rm d} \approx R_{\rm e}/1.67$ \citep{Binney&Tremaine87}, where $R_{\rm e}$ is the effective radius.} This choice is motivated by the fact that a self-gravitating exponential disk reaches its maximum rotation velocity at $\sim 2.2R_{\rm d}$ \citep{Freeman70}. We note that this rotation velocity definition (often denoted $V_{2.2}$) is commonly adopted in H$\alpha$-based Tully-Fisher studies (e.g., \citealt{Courteau97}; \citealt{Reyes+11}). 

{ We measure the global velocity dispersion of each galaxy from the corresponding aperture spectrum, correcting the observed line width for instrumental broadening using the LSF. We adopt an extraction radius of $2.2R_{\rm d}$ as a compromise between S/N and beam-smearing effects (e.g., \citealt{Wisnioski+15,Johnson+18,Puglisi+23}). That is, we avoid extracting values from the central regions where velocity gradients are strongest.} We present the kinematic parameters for our galaxy disks in Table~2.

\subsection{Asymmetric drift correction and circular velocity}
Rotational velocities measured from ionized gas kinematics can be affected by pressure support, especially in high-$z$ galaxies where intrinsic velocity dispersions are high \citep{Forster-schreiber&wuyts20}. This effect reduces the gas rotation velocity, so an asymmetric drift correction must be applied to $V_{2.2}$ to estimate $V_{\rm circ}$. 

We follow the approach of \cite{Burkert+10}, where the circular velocity can be estimated as
\begin{equation}
    \label{eq:ADC}
    V_{\rm circ}^2(R) = V_{\rm rot}^2(R) + 2\,\sigma_0^2 \left( \frac{R}{R_{\rm d}} \right),
\end{equation}

\noindent where $R$ is the radius at which $V_{\rm rot}$ is derived. This asymmetric drift correction assumes: (i) a constant intrinsic velocity dispersion across the disk, (ii) an exponential surface mass density profile, and (iii) isotropic turbulent pressure support. We list the $V_{\rm circ}$ values in Table~\ref{tab2}.

\begin{table*}
    \def\arraystretch{1.3}
    \setlength\tabcolsep{3pt}
    \caption{Kinematic and morphological properties of the disk galaxies}
    \vspace{-0.2mm}
    \label{tab2}
    \centering
    \small
    \renewcommand{\arraystretch}{1.5}
    \begin{tabular}{lccccccccccc}
        \hline
        ID & { $z_{\rm H\alpha}$} & \multicolumn{2}{c}{$R_e$} 
        & $i$ & { PA$_{\rm kin}$} & $V_{2.2}$ 
        & { $V_{2.2}^{\rm DYSMALpy}$}
        & $\sigma_0$ 
        & { $\sigma_0^{\rm DYSMALpy}$} 
        & $V_{\rm circ}$ 
        & { $V_{\rm circ}^{\rm DYSMALpy}$} \\
        
        & & ($''$) & { (kpc)} & ($^\circ$) & ($^\circ$) 
        & (km\,s$^{-1}$)
        & { (km\,s$^{-1}$)}
        & (km\,s$^{-1}$) 
        & { (km\,s$^{-1}$)}
        & (km\,s$^{-1}$) 
        & { (km\,s$^{-1}$)} \\
        
        (1) & { (2)} & (3) & { (4)} & (5) & (6) 
        & (7) & { (8)} & (9) & { (10)} & (11) 
        & { (12)} \\ 
        \hline
        
        DC--494763 
        & ${ 5.2337}$
        & $0.19$ & ${ 1.17}$ & $53 \pm 2$ & $108 \pm 7$ 
        & $109 \pm 4$
        & ${ 127 \pm 4}$
        & $\phantom{0}51 \pm 20$ 
        & ${ 66 \pm 13}$
        & $153 \pm 29$ 
        & ${ 186 \pm 20}$ \\

        DC--519281 
        & ${ 5.5759}$
        & $0.16$ & ${ 0.95}$ & $45 \pm 3$ & $319 \pm 9$ 
        & $\phantom{0}97 \pm 5$
        & ${ 113 \pm 5}$
        & $126 \pm 39$ 
        & ${ 129 \pm 26}$
        & $282 \pm 77$ 
        & ${ 315 \pm 55}$ \\

        DC--536534 
        & ${ 5.6886}$
        & $0.21$ & ${ 1.23}$ & $57 \pm 2$ & $333 \pm 3$ 
        & $\phantom{0}80 \pm 4$
        & ${ 113 \pm 3}$
        & $115 \pm 39$ 
        & ${ 113 \pm 23}$
        & $254 \pm 78$ 
        & ${ 219 \pm 32}$ \\

        DC--630594 
        & ${ 4.4403}$
        & $0.22$ & ${ 1.46}$ & $65 \pm 2$ & $143 \pm 6$ 
        & $\phantom{0}65 \pm 6$
        & ${ 110 \pm 9}$
        & $\phantom{0}80 \pm 29$ 
        & ${ 78 \pm 16}$
        & $179 \pm 56$ 
        & ${ 195 \pm 27}$ \\

        DC--742174 
        & ${ 5.6360}$
        & $0.40$ & ${ 2.36}$ & $76 \pm 7$ & $\phantom{0}87 \pm 10$ 
        & $\phantom{00}90 \pm 27$
        & ${ 91 \pm 3}$
        & $84 \pm 46$ 
        & ${ 82 \pm 16}$
        & $197 \pm 105$ 
        & ${ 197 \pm 31}$ \\ 
        
        \hline
    \end{tabular}
    \renewcommand{\arraystretch}{1.0}
    
    \tablefoot{
    (1) Source name. 
    { (2) Redshift measured from the H$\alpha$ emission line.}
    (3--4) Effective radius measured from the JWST/F444W. 
    The average error computed by {\sc pysersic} is $<0\farcs01$. 
    (5) Inclination angle. 
    (6) Kinematic position angle, measured counterclockwise from north 
    to the receding side of the velocity map. 
    (7) Inclination-corrected rotation velocity measured at 
    $2.2\,R_{\rm d}$ using our non-parametric approach. 
    { (8) Intrinsic rotation velocity at $2.2\,R_{\rm d}$ 
    inferred by {\sc DYSMALpy}.}
    (9) Non-parametric velocity dispersion measurement at $2.2R_{\rm d}$. 
    { (10) Intrinsic velocity dispersion inferred by {\sc DYSMALpy}.} 
    (11) Circular velocity at $2.2\,R_{\rm d}$ estimated from the 
    non-parametric approach. 
    { (12) Circular velocity at $2.2\,R_{\rm d}$ inferred from the 
    best-fitting {\sc DYSMALpy} parameters.}
    }
\end{table*}

\subsection{Forward modeling with {\sc DYSMALpy}}
\label{sec:dysmalpy}
The limited spectral and spatial resolution of the kinematic data of high-$z$ galaxies imposes severe limitations when recovering their intrinsic kinematic properties. Different kinematic routines have been developed to overcome such limitations, but systematic uncertainties inherent in the assumptions made by these tools can still bias the kinematic measurements (e.g., \citealt{Lee+25,Yttergren+25}). { To address the effects of observational systematics on our kinematic parameters, we use the forward-modeling code {\sc DYSMALpy} \citep{Price+21,Lee+2025} to derive the intrinsic kinematic parameters. The main advantage of {\sc DYSMALpy} is that projection effects and observational broadening from the PSF and LSF are included directly in the forward modeling.}

{\sc DYSMALpy} fits composite mass models by generating mock kinematic observables, which can be a data cube, two-dimensional maps, or a rotation curve and a velocity dispersion profile, and compares them to the observed data. { In this work, we fit the one-dimensional rotation curve and velocity dispersion radial profiles derived in Section~\ref{subsec:kin_pars}.} The composite mass model can include a bulge, a disk, and a dark-matter halo. The baryonic mass component, bulge and disk, corresponds to the sum of two S\'ersic profiles, where the bulge prominence is parameterized by the bulge-to-total mass ratio ($B/T$). However, since prominent bulges are not usual in typical star-forming galaxies at $z = 4-6$ \citep{Ferreira+23,Huertas-Company+24}, and we do not observe any clear stellar features that could be regarded as bulges in the JWST/NIRCam F444W images, we set $B/T=0$. We also set the intrinsic thickness of the disk equal to 0.2. We further assume a radially uniform intrinsic velocity dispersion, parameterized by a single value $\sigma_0$. For the dark-matter halo, we assume a two-parameter Navarro-Frenk-White profile \citep{Navarro+96}. { We parameterize the halo contribution through the dark matter fraction at the effective radius, defined as $f_{\rm DM}(R_e)=V_{\rm DM}^2(R_e)/V_{\rm circ}^2(R_e)$. For each trial value of $f_{\rm DM}(R_e)$, {\sc DYSMALpy} determines the corresponding NFW virial mass given the baryonic mass distribution and the adopted halo concentration.} { For each galaxy, we use the stellar-to-halo mass relation of \citet{Girelli+20} to define a broad halo-mass range consistent with its stellar mass, adopting $M_{\rm vir}=10^{11}-10^{13}\,M_\odot$.}
We fix the concentration parameter ($c$) to $3.5$, following the $M_{\rm vir}-c$ relation of \citet{Dutton&Maccio14}. We compute the circular velocity by adopting the asymmetric drift approximation of \cite{Noordermeer08} within {\sc DYSMALpy}, so that the correction factor is calculated in the same way as described in Equation~\ref{eq:ADC}. The results obtained with {\sc DYSMALpy} are presented in Table~\ref{tab2}.

\section{Results}
\label{sec:results}

\subsection{Kinematic properties of the ALPINE-CRISTAL-JWST sample}
\label{sec:kin_properties}

\begin{figure*}
    \centering
    \includegraphics[width=1\linewidth]{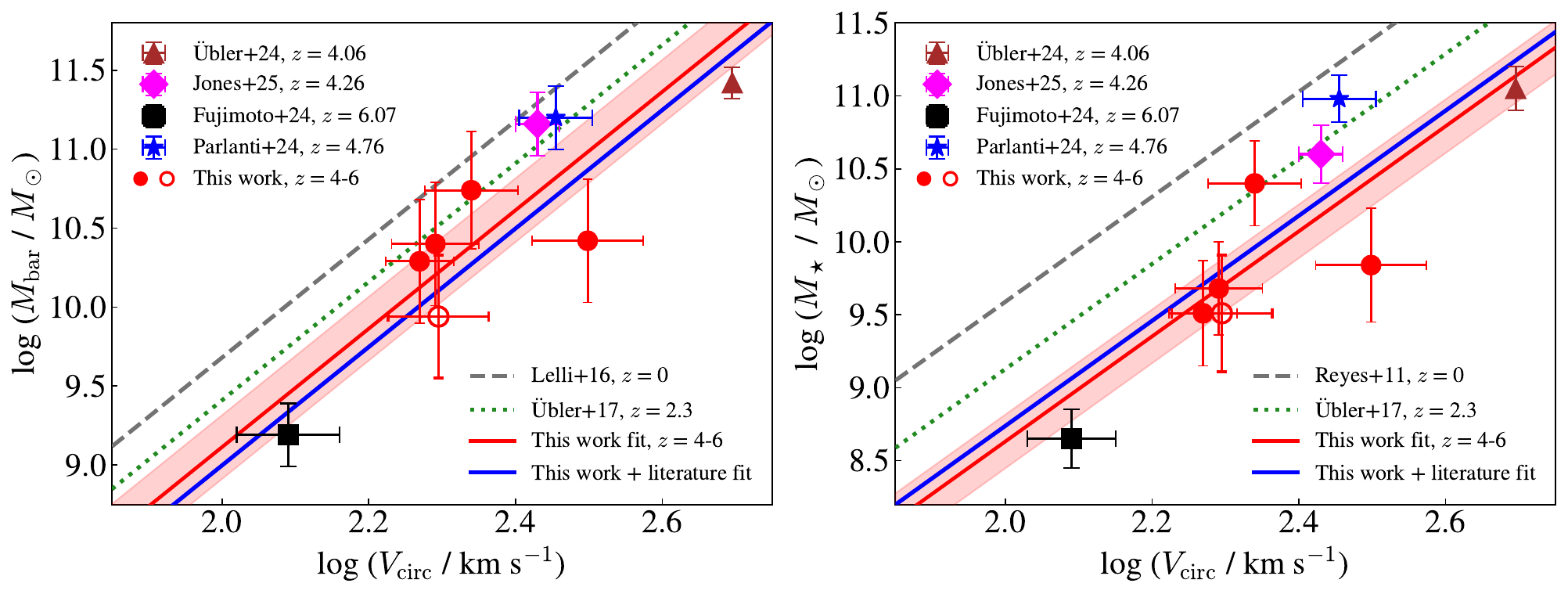}
\caption{Left: Baryonic-mass Tully-Fisher relation. Right: Stellar-mass Tully-Fisher relation. In both panels, the best-fit relations are obtained by fixing the slope and fitting only the zero-point. The red-shaded regions show the $1\sigma$ uncertainties of the fits based only on the data presented in this work. The literature data correspond to galaxies at similar redshifts with H$\alpha$-based kinematic measurements provided by NIRSpec/IFU observations. { In both panels, the open circle represents DC--742174 (Section~\ref{subsec:disk_class}).}}

    \label{fig:TFR}
\end{figure*}

Figure~\ref{fig:summary_5x5} presents the H$\alpha$ moment maps, rotation curves, and velocity dispersion radial profiles for the disk sample. The H$\alpha$ moment maps for the full sample are shown in Appendix~\ref{app:allsample}. The $V_{\rm LOS}$ maps show the typical blue-red pattern that indicates rotation. In all cases, the rotation curves and velocity dispersion radial profiles extend beyond $2.2\,R_{\rm d}$, allowing us to estimate { $V_{2.2}$} and $\sigma_0$. For four galaxies, our inferred $\sigma_0$ estimates are lower than the instrumental resolution value ($\sim 127$\,km\,s$^{-1}$). This is because our emission-line modeling includes the LSF (Section~\ref{sec:methods}) and the fitted emission lines have S/N high enough so that they are detected over several consecutive channels. We verify this by fitting mock spectra specifically designed to mimic the NIRSpec/IFU medium-filter spectral resolution. We find that intrinsic $\sigma_0$ values $\sim 60-80$\,km\,s$^{-1}$ can be recovered with typical errors below 40\% if the fitted line has a $S/N \gtrsim 10$ (see Appendix~\ref{app:vsig_sim} for more details). { For the non-parametric measurements, the uncertainties inferred from these tests are included in the $\sigma_0$ error budget and propagated to the corresponding circular velocities.}

{ The non-parametric measured kinematics described in section~\ref{subsec:kin_pars} can still be affected by beam smearing, even at the relatively large radii considered here. We therefore compare the non-parametric measurements with the intrinsic kinematic parameters inferred from the {\sc DYSMALpy} forward models, which explicitly account for projection effects and convolution with the spatial PSF and spectral LSF. The intrinsic $V_{2.2}$ values inferred with {\sc DYSMALpy} are systematically larger than the directly measured inclination-corrected values, by a factor of $\sim1.3$ on average ($\sim0.10$ dex). This shows that the limited spatial resolution suppresses the observed rotation velocities. In contrast, the intrinsic $\sigma_0$ values differ from the non-parametric measurements by only $\sim0.02$ dex on average, consistent within their uncertainties.}

{ As independent checks, we also applied the empirical beam-smearing corrections of \citet{Johnson+18} and performed independent 3D forward modelling with {\sc 3DBarolo} \citep{DiTeodoro+15}. Both methods imply smaller corrections to $V_{2.2}$, typically by a factor of $\sim1.1$, compared with the $\sim1.3$ correction inferred with {\sc DYSMALpy}. These tests show that the observed rotation velocities are affected by the limited spatial resolution.}

{ For the remainder of the analysis, we adopt the intrinsic kinematic parameters inferred with {\sc DYSMALpy} as our fiducial values. Importantly, the differences among these beam-smearing treatments do not significantly affect our main results as discussed below.}

{ Using our fiducial {\sc DYSMALpy} measurements, the disk-like sample spans $V_{2.2}=91-127$\,km\,s$^{-1}$ and $\sigma_0=66-129$\,km\,s$^{-1}$, corresponding to $V_{2.2}/\sigma_0\simeq0.9-1.9$.} These values indicate large random motions and significant asymmetric-drift corrections to be applied for estimating $V_{\rm circ}$. We discuss the impact of the adopted asymmetric-drift prescription on our results in Appendix~\ref{app:adc}. { After accounting for pressure support, we infer $V_{\rm circ}=186-315$\,km\,s$^{-1}$ for the disk-like sample (Table~\ref{tab2}).}

\subsection{The stellar-mass Tully-Fisher relation}

Figure~\ref{fig:TFR} presents our targets in the context of the stellar-mass Tully-Fisher relation (sTFR). As a reference, we also show the local sTFR of \citet{Reyes+11} and the sTFR at $z \approx 2.3$ taken from \citet{Ubler+17}.\footnote{\citet{Reyes+11} define the local TFR relation using the outer rotation velocity $V_{80}$. \citet{Reyes+11} show that $V_{80}$ typically differs by only a few percent from $V_{2.2}$ for most galaxies. Thus, we consider $V_{80}$ and $V_{2.2}$ as comparable velocity measures for disk systems. We also note that our data analysis is consistent with that of \citet{Ubler+17}.} We find that our disk galaxies are clearly offset from the local and cosmic-noon sTFRs.\footnote{{ \citet{Ubler+17} adopt SED-based stellar masses, while \citet{Reyes+11} use colour-based mass-to-light ratios, which can differ from SED-based estimates by $\sim0.15$ dex.}}

To quantify the average offset of our measurements with respect to the low-$z$ sTFRs, we fit a linear function of the form

\begin{equation}
\log\left(\frac{M_\star}{{M}_\odot}\right) = \alpha  \log\left(\frac{V_{\rm circ}}{\rm km,s^{-1}}\right) + \beta.
\end{equation}

\noindent where $\alpha$ is the slope and $\beta$ is the zero-point. Due to the small sample size, we cannot robustly constrain the slope. Therefore, we set $\alpha = 3.6$, the local universe value \citep{Reyes+11}, and fit only for the zero-point $\beta$, as commonly done in high-$z$ TFR studies (e.g., \citealt{Puech+08,Tiley+16,Ubler+17,Sharma+24}). We perform the fit using {\sc lmfit}, obtaining { $\beta = 1.43 \pm 0.18$}. This value corresponds to a zero-point offset of { $-0.96 \pm 0.18$\,dex} with respect to the local sTFR of \citet{Reyes+11} and { $-0.50 \pm 0.18$\,dex} lower relative to the $z\approx 2.3$ sTFR of \citet{Ubler+17}. { Varying $q_0$ from 0.2 to 0.5 changes the sTFR zero-point offset by 0.03\,dex, below its statistical uncertainty. A similar variation of the sTFR zero-point is found when removing DC--742174 from the analysis.} 

We searched the literature for systems at $z=4-6$ with JWST/NIRSpec IFU data delivering H$\alpha$ kinematics. We find four rotating or disk-like galaxies: a low-mass clumpy galaxy at $z \approx 6$ with $M_\star = 4.5^{+2.7}_{-1.1}\times10^8 , {\rm M}_\odot$ and $\mathrm{SFR} = 2.6^{+1.7}_{-1.1} ,{\rm M}_\odot,\mathrm{yr}^{-1}$ \citep{Fujimoto+25a}; GN20, an AGN host galaxy at $z \approx 4$ with $M_\star = 1.1 \times10^{11} , {\rm M}_\odot$ \citep{Ubler+24}; a rotating galaxy at $z = 4.26$ with $M_\star \approx 4 \times10^{10} , {\rm M}_\odot$ and $\mathrm{SFR} = 59 \pm 5 , {\rm M}_\odot,\mathrm{yr}^{-1}$ \citep{Jones+25} and another AGN host galaxy at $z = 4.76$ with M$_\star = 10^{10.98} , {\rm M}_\odot$ \citep{Parlanti+24}. { When we include these systems in the $M_\star-V_{\rm circ}$ plane and re-fit the relation, we find $\beta = 1.54 \pm 0.09$, corresponding to a zero-point offset of $-0.85 \pm 0.09$\,dex with respect to the local sTFR of \citet{Reyes+11} and $-0.39 \pm 0.09$\,dex relative to the $z\approx2.3$ sTFR of \citet{Ubler+17}. The fitted zero-point changes by only $\Delta\beta = +0.11$\,dex relative to the fit to our sample alone, which is smaller than the $1\sigma$ uncertainty of the latter.} { By using {\sc pymc}, we estimate a sTFR intrinsic scatter of $0.29$\,dex. We note, this value is not particularly driven by the inclusion of any of our targets or the literature data.}

\subsection{The baryonic-mass Tully-Fisher relation}

In the local Universe, gas-rich low-mass galaxies are outliers of the sTFR. When the gas component is included in the mass budget, these galaxies { follow} a relation in terms of baryonic mass ($M_{\rm bar}=M_\star+M_{\rm gas}$), further extending the sTFR in what has been called the baryonic-mass TFR (bTFR; \citealt{Mcgaugh+00,Lelli+16}). Galaxies at high-$z$ are commonly gas rich \citep{Tacconi+20}, suggesting that studying the bTFR may be better to determine any offset of the TFR zero-point.

To estimate the cold-gas mass for our galaxies, we take the \cii luminosities presented for the ALPINE galaxies \citep{Bethermin+20} and adopt the $L_{\rm [CII]}-M_{\rm gas}$ prescription of \citet{Zanella+18}, with a scatter of $\sim 0.3$\,dex. { We include this scatter in quadrature in the $M_{\rm bar}$ uncertainty}. On average, we find that our galaxies are gas rich, with a mean sample gas fraction ($f_{\rm gas} = M_{\rm gas}/M_{\rm bar}$) of $0.68 \pm 0.10$. This average gas fraction is consistent with the expected evolution of gas content in main-sequence galaxies with $M_\star \approx 10^{10},{\rm M}_\odot$ at $z \approx 5$ \citep{Tacconi+20}, and with the mean $f_{\rm gas}$ value of the ALPINE parent sample ($f_{\rm gas} = 0.63 \pm 0.03$, \citealt{Dessauges-Zavadsky+20}). These \cii-based gas masses are also consistent with the dust-based estimates of \citet{Lee+25} and with those obtained using the metallicity-dependent \cii prescription of \citet{Vallini+25}\footnote{We adopt the gas-phase metallicities provided by \cite{Faisst+26b} for our sample.}. Relative to the \citet{Zanella+18} estimates, { we find mean offsets of $-0.26$ dex and $-0.18$ dex for \citet{Lee+25} and \citet{Vallini+25}, respectively. Both offsets are consistent within the adopted $\sim0.3$ dex uncertainty of the \citet{Zanella+18} conversion.}

Figure~\ref{fig:TFR} shows the bTFR for our disk galaxy sample. As a reference, we also show the local universe bTFR of \citet{Lelli+16} and the bTFR at $z \approx 2.3$ taken from \citet{Ubler+17}. We find all our sources located below the local bTFR. Three sources closely follow the cosmic noon bTFR, while the other two systems are offset from this relation. By fitting a linear function, keeping the slope fixed to the local value ($\alpha = 3.75$; \citealt{Lelli+16}), we find a zero-point offset of { $-0.57 \pm 0.20$\,dex} relative to the local bTFR, and { $-0.30 \pm 0.20$\,dex} relative to the $z\approx 2.3$ bTFR. In both cases, the zero-point offsets are smaller compared to those measured for the sTFR. Similarly to the sTFR case, the literature data broadly agree with our sample. { When we include these data in the fit, we obtain $\beta = 1.50 \pm 0.08$, corresponding to a zero-point offset of $-0.69 \pm 0.08$ dex with respect to the local bTFR of \citet{Lelli+16} and $-0.42 \pm 0.08$ dex relative to the $z\approx2.3$ bTFR of \citet{Ubler+17}. The fitted zero-point changes by $\Delta\beta=-0.12$ dex relative to the fit to our sample alone, which is smaller than the $1\sigma$ uncertainty of the latter. We note that, varying $q_0$ from 0.2 to 0.5 changes the inferred bTFR zero-point offset by 0.03 dex, and removing DC--742174 from our analysis changes the inferred zero-point offset by 0.07 dex. In both cases, the bTFR zero-point variation is well below the $1\sigma$ uncertainty.}

{ We also perform a fit to the bTFR considering cold-gas masses inferred from the metallicity-dependent \cii prescription of \citet{Vallini+25} and from the dust-based molecular gas fractions of \citet{Lee+25}. For a consistent comparison, we use the four galaxies for which gas masses are available with all three prescriptions and compute the zero-point variation relative to that of assuming the conversion factor of \citep{Zanella+18}. The zero-point varies by $\Delta\beta=-0.14\pm0.22$\,dex and $-0.12\pm0.34$\,dex when adopting the \citet{Vallini+25} and \citet{Lee+25} prescriptions, respectively. These variations imply that our inferred bTFR zero-point offset is not driven by the specific cold-gas mass prescription adopted.} { By using {\sc pymc}, we estimate a bTFR intrinsic scatter of $0.28$\,dex.}

\section{Discussion}
\label{sec:discussion}

\subsection{Evolution of the TFR with cosmic time}
\label{sec:TFR_evol}

\begin{figure*}
    \centering
    \includegraphics[width=1\linewidth]{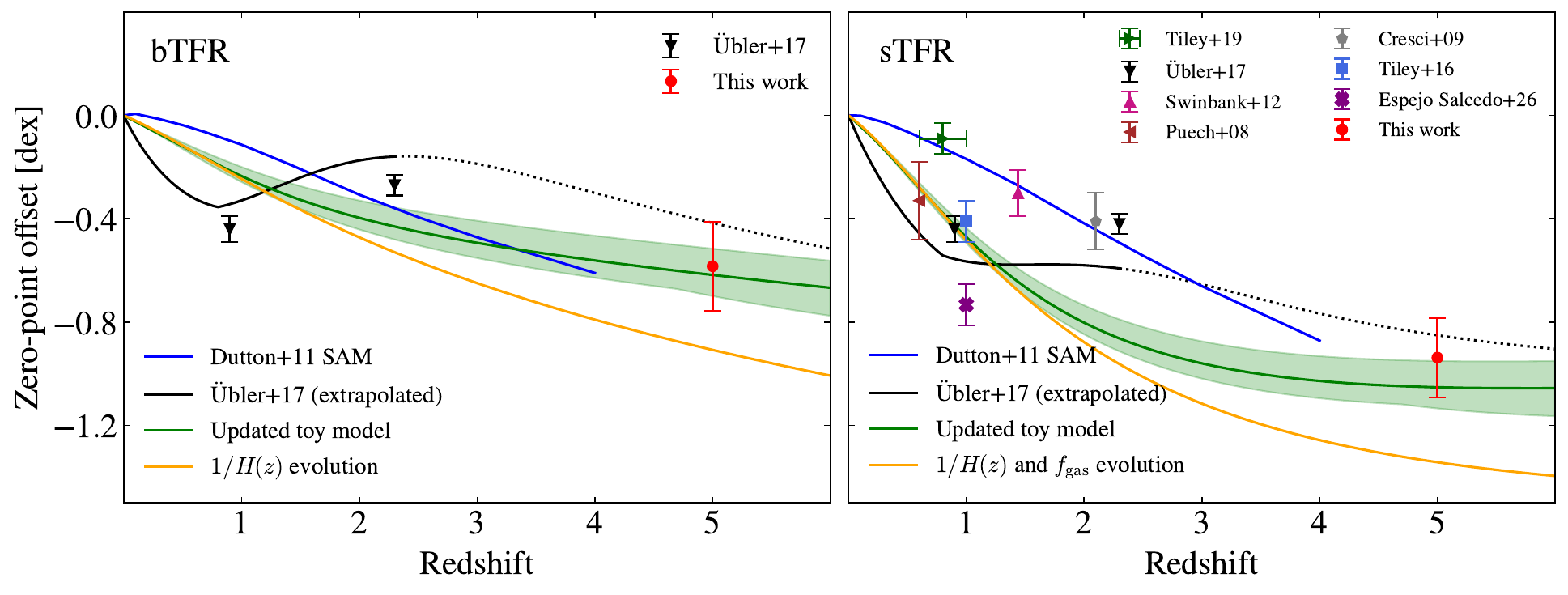}
    \caption{{\it Left}: Baryonic-mass TFR zero-point offset as a function of redshift. The error bars show the $1\sigma$ uncertainty. The green curve shows the model of \citet{Ubler+17} updated by adopting the \cite{Tacconi+20} gas fraction evolution, \cite{Girelli+20} stellar-to-halo mass relation, and an exponential fit for the cosmic evolution of the dark matter fraction within galaxies (see Section~\ref{sec:TFR_evol}). The green-shaded region shows the variation of the updated  model subject to a $f_{\rm DM}$ value change of $\pm 0.1$. The dotted lines show an extrapolation of the original model of \citet{Ubler+17} up to $z=6$. {\it Right}: Stellar-mass TFR zero-point offset as a function of redshift. We color-code as the same as the left panel.}

    \label{fig:zp_evolution}
\end{figure*}

Our study provides the first glimpse into the properties of disk galaxies and the Tully-Fisher relation at $z \approx 5$. Our findings suggest a negative variation of the TFR zero point with redshift for both baryonic and stellar TFRs, meaning that galaxies at $z \approx 5$ appear to have higher circular velocities for a fixed stellar or baryonic mass (Figure~\ref{fig:TFR}). Comparisons with galaxy samples at lower redshifts are not straightforward as studies { have reported both weak and significant evolution of the sTFR \citep[e.g.][]{Puech+08,Vergani+12,tiley+19,Espejo-Salcedo+25,Mancera-Pina+26}.} For example, \citet{Ubler+17} suggested a negative zero-point evolution up to $z \approx 0.9$ for both bTFR and sTFR, and no evolution of the sTFR from $z \approx 0.9$ to 2.3, which contrasts with the negative evolution of the bTFR reported over the same redshift range. Conversely, \citet{tiley+19} suggest only a mild negative evolution of the sTFR at $z \approx 1$, consistent with \citet{Miller+11,Miller+12} and the predictions of N-body simulations of baryonic disks growing in dark matter halos of \citet{Dutton+11}. More recently, \citet{Sharma+24} reported a subtle evolution of the zero points of the bTFR and sTFR up to $z \sim 0.6-2.5$. These contrasting results likely reflect differences in sample selection and/or kinematic modeling strategies when estimating $V_{\rm circ}$.

Interpreting the cosmic evolution of the zero points of the bTFR and sTFR is not straightforward. In the framework of virialized dark matter halos, the cosmic evolution of the zero point of the bTFR reflects the effect of smaller DM halo sizes at high-$z$, if the baryonic-to-halo mass ratio ($m_d \equiv M_{\rm bar}/M_{\rm halo}$) { is assumed to remain constant with redshift}. Because DM halo masses scale inversely with the Hubble parameter, $H(z)$, a smooth evolution is expected. This contrasts with the variation of the zero point of the sTFR with redshift, which is more convoluted due to the cosmic evolution of the $f_{\rm gas}$ \citep{Tacconi+20}, which also plays a role \citep{Ubler+17}. For both bTFR and sTFR, further considerations regarding the internal mass distribution of galaxies must be made because $V_{\rm circ}$ is usually estimated at inner radii (e.g., $\approx 2.2 \, R_{\rm d}$), not at the virial radius of the halo. The low DM fraction measured in the central regions of galaxies at high-$z$ \citep{Genzel+17,Lang+17,Genzel+20} biases the central $V_{\rm circ}$ to low values, shifting galaxies leftward in the $M_{\rm bar}-V_{\rm circ}$ plane, such that any possible evolution of the TFR could be erased or even reversed \citep{Ubler+17}. To account for these effects, we compare our measurements with the predictions of the empirical model of \citet{Ubler+17}:

\begin{subequations}
 \begin{align}
    M_{\rm bar} &\propto  \frac{V_{\rm circ}^3\, \left[1-f_{\rm DM}(<R_{\rm e},z)\right]^{3/2}\,}{H(z)\,\left[m_{d}(z)\right]^{1/2}} \label{eq:toy_ubler}, \\
    M_\star &\propto \frac{V_{\rm circ}^3\,\left[1-f_{\rm DM}(<R_{\rm e},z)\right]^{3/2}\, \left[1-f_{\rm gas}(z)\right]}{H(z)\,\left[m_{d}(z)\right]^{1/2}}, \label{eq:toy_stfr_ubler}
 \end{align}
\end{subequations}

\noindent where $f_{\rm DM}(<R_{\rm e},z)$ is the dark matter fraction within one effective radius. The model assumes virialized DM halos, { galaxy sizes proportional to their parent halo sizes, and baryonic masses related to those of the halos through $f_{\rm DM}$ (see Appendix D of \citealt{Ubler+17} for its full derivation, and \citealt{Mancera-Pina+26} for a similar model). In this toy model, the} variation of the TFR zero point with redshift can be further affected by (i) changes in the baryonic-to-halo mass ratio, (ii) the evolution of the DM fraction in the inner regions of the halo (parameterized at the optical radius), and (iii) changes in the gas fraction for the sTFR. Initially, this model was introduced to study the evolution of the TFR at cosmic noon by \citet{Ubler+17}. Here, we update it by adopting more recent parameterizations of $f_{\rm DM}(R_e, z)$, $f_{\rm gas}(z)$, and $m_d(z)$ for main-sequence galaxies with $M_\star=10^{10}\,{\rm M}_\odot$ and $z \lesssim 5$. We adopt the parameterization for $f_{\rm gas}(z)$ from \citet{Tacconi+20}. To derive $m_d(z)$, we assume a stellar-to-halo mass ratio combined with $f_{\rm gas}(z)$. We assume a constant stellar-to-halo mass ratio of $0.020 \pm 0.005$ due to its expected modest cosmic evolution for galaxies with $M_\star=10^{10}\,{\rm M}_\odot$ \citep{Girelli+20}. To compute $f_{\rm DM}(R_{\rm e},z)$, we combine measurements from the literature and interpolate them with redshift. We adopt $f_{\rm DM}(<R_{\rm e})\approx0.64$ for galaxies at $z \approx 0$ \citep{Martinsson+13}. At $0.6\lesssim z\lesssim2.5$, we consider $f_{\rm DM}(R_{\rm e},z)$ values taken from the RC100 sample \citep{Nestor-Shachar+23}. Finally, for systems at $z\approx5$, we include the average estimate for CRISTAL galaxies \citep{Lee+25}. We interpolate these values by fitting a smooth redshift function, obtaining $f_{\rm DM}(z) = 0.58 \, \exp[(-0.31 \times z)^{1.5}]$ (see Appendix~\ref{app:fdm} for more details).

Figure~\ref{fig:zp_evolution} shows the zero-point variation with redshift for the bTFR and sTFR. As a reference, we also include the expected trends for virialized halos in the bTFR, scaling as the inverse of $H(z)$, and its conversion to the sTFR when accounting for the cosmic evolution of $f_{\rm gas}$. In both cases, these predicted trends overestimate the zero-point offsets measured by $\sim -0.3\,$dex at $z \approx 5$, unlike the empirical model, which remains consistent within $1\sigma$ uncertainty. The updated and the original empirical model of \cite{Ubler+17} predict different variations in the bTFR zero-point offset with redshift, highlighting how sensitive these are to the adopted $f_{\rm DM}(<R_e,z)$ prescription (Appendix~\ref{app:fdm}). However, in both model versions, the relatively small cosmic evolution of the bTFR (and sTFR) is due to high-$z$ massive galaxies presenting low DM fractions in their inner regions.

Our empirical model does not include any variation in the intrinsic properties of the DM halos (e.g., concentration, spin) with redshift \citep{Ubler+17}. Such variations could occur depending on how baryons are accreted and assembled in the inner halo regions \citep{Glowacki+21,Sorini+25}. For example, decreasing halo concentrations with redshift could reduce the cosmic evolution of the bTFR zero point, as highlighted by the semi-analytic model (SAM) of \cite{Dutton+11}. Although that model predicts zero-point offsets only up to $z \approx 4$ for the bTFR and sTFR, these trends broadly agree with the empirical model of \citet{Ubler+17} for $z \gtrsim 2$ systems, and are also broadly consistent with our results.

We caution that the inferred zero-point offsets are based on a small number of disk galaxies. Although fixing the slope reduces the number of free parameters, small samples can still introduce statistical fluctuations in the recovered TFR normalization. To estimate this effect, we performed recovery tests by generating mock linear TFRs with a known input offset and re-fitting randomly selected subsamples of five and nine galaxies. On average, the input offsets were recovered without significant bias, with a mean difference between the recovered and input zero-points of $\sim 0.01$\,dex. The scatter in the recovered offsets was $\sim 0.25$\,dex for five galaxies and $\sim 0.18$\,dex for nine galaxies, both smaller than the offsets reported here. This suggests that the measured negative evolution is unlikely to be driven entirely by small-number statistics. Nevertheless, the exact amplitude of the evolution should be regarded as tentative until larger samples of spatially resolved disk galaxies become available.

\subsection{Are we witnessing the settling of the first galactic disks?}

\begin{figure}
    \centering
    \includegraphics[width=0.95\linewidth]{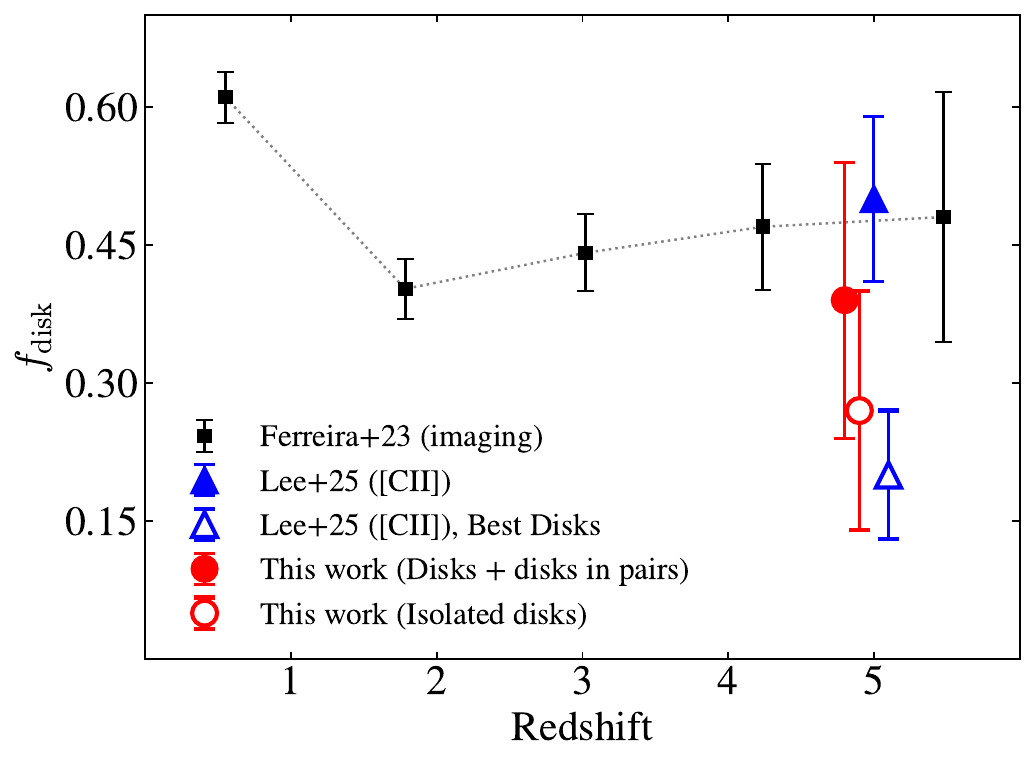}
    \caption{Cosmic evolution of the disk galaxy fraction. We report both values with and without considering the disks in galaxy pairs in our sample. We slightly shift in redshift the data taken from \cite{Lee+25} to improve visualization.}
    \label{fig:disks_fractions}
\end{figure}

\begin{figure*}
    \centering
    \includegraphics[width=1\linewidth]{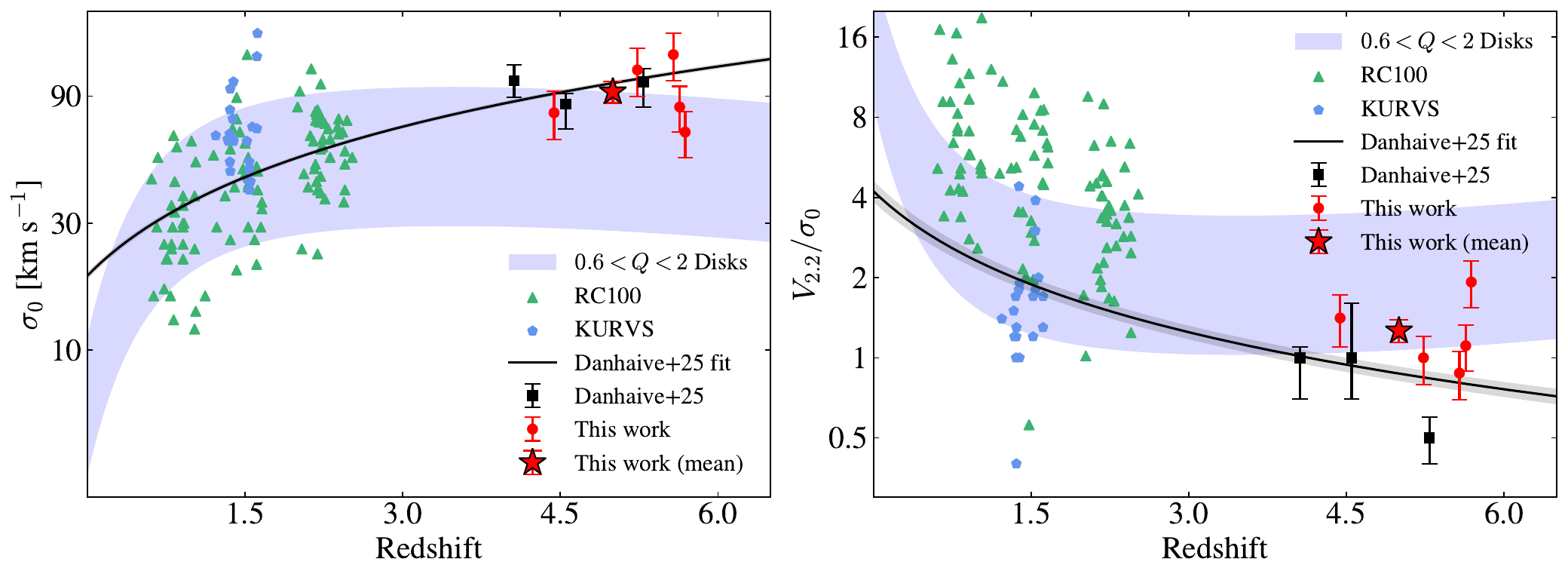}
    \caption{{\it Left}: Cosmic evolution of the H$\alpha$-based intrinsic velocity dispersion of star-forming galaxies. {\it Right}: Rotation-to-velocity dispersion ratio as a function of redshift. In both panels, the shaded region corresponds to the evolution of $\sigma_0$ or $V_{\rm rot}/\sigma_0$ predicted for marginally stable disks (e.g., \citealt{Wisnioski+15}). Green and blue points represent H$\alpha$-based values drawn from the RC100 \citep{Nestor-Shachar+23} and KURVS \citep{Puglisi+23} surveys. The data taken from \cite{Danhaive+25} correspond to the median values of their sample.}
    \label{fig:disks}
\end{figure*}

Recent studies of galaxy morphology \citep{Ferreira+23,Kartaltepe+23,Huertas-Company+24} and cold-gas kinematics \citep{Lee+25} report that $\sim 30\%-50\%$ of the $z \approx 5$ systems resemble disk-like structures. We also find a comparable disk fraction ($f_{\rm disk}$) in our sample, with five disk-like systems out of 18, corresponding to $f_{\rm disk}=0.28\pm0.11$.\footnote{We estimate the $1\sigma$ uncertainty assuming binomial statistics.} If we also consider galaxies in pairs { which fulfill our disk criteria}, we find a total of eight disk-like systems, yielding $f_{\rm disk} = 0.36\pm0.10$. (Figure~\ref{fig:disks_fractions}). These values indicate that a non-negligible subset of massive galaxies ($M_\star \gtrsim 10^{9.5}$) has already developed a rotating component by $z\approx5$, while a larger fraction of the population still exhibits disturbed kinematics and/or is undergoing merging. 

It remains unclear how these disk galaxies arise in the early universe and the main mechanisms driving their evolution. On one hand, frequent interactions can perturb and disrupt forming disks, so observations of both isolated and interacting disks can be interpreted as galaxy growth occurring in environments where major encounters remain common \citep{Rodriguez-Gomez+15,Puskas+25,Duan+25}. On the other hand, high gas accretion rates and strong internal gravitational instabilities can lead to increased turbulence within the disks, keeping them evolving in a gravitational state marginally stable (e.g., \citealt{Wisnioski+15,Stott+16}). The average velocity dispersion ($\sim 90$\,km\,s$^{-1}$) and $V_{2.2}/\sigma_0$ ratio ($= 1.26 \pm 0.19$) measured for our disk galaxies are broadly consistent with the latter scenario (Figure~\ref{fig:disks}), as they approximately follow the cosmic evolutionary trend predicted by the marginally stable disk model of \citet{Wisnioski+15}, which assumes a \citet{Toomre64} parameter $Q \approx 1$ { (see \citealt{Bacchini+24} for some limitations of this model). Even though we do not directly compute $Q$, the rough agreement with the cosmic evolutionary trend suggest that the} kinematics of these galactic disks at $z\approx 5$ likely reflect the balance between rotation, pressure support, and feedback-driven turbulence as they grow and settle while their gas supply changes \citep{Kassin+12,Wisnioski+15,Simons+17}.

The kinematic tracer we used, H$\alpha$, could include some contribution from { non-circular motions, such as} ionized gas outflow or extraplanar regions. As a result, intrinsic velocity dispersion from ionized gas may overestimate the disk mid-plane value, which can be more reliably recovered from cold-gas kinematics \citep{wisnioski+25}. However, comparisons with available \cii kinematic data from the CRISTAL survey suggest that this is not the case. For the four disk-like galaxies in common with \citet{Lee+25}, { we find that their average intrinsic cold-gas velocity dispersion is $\sim87$\,km\,s$^{-1}$, consistent with our H$\alpha$-based measurements ($\sim 93$\,km\,s$^{-1}$). The full CRISTAL disk sample has a median intrinsic dispersion of $\sim70$\,km\,s$^{-1}$, consistent with the high gas fractions of systems and turbulence being driven by gravitational instabilities and gas transport \citep{Lee+25}}. This is somewhat different from the predictions of post-processed simulations \citep{Kohandel+24}, which suggest a factor of $\sim 2$ difference between the velocity dispersion of the ionized and molecular gas ISM phases for the stellar-mass range probed by our disk galaxy sample ($M_\star \approx 10^{9.5-10.5}\,{\rm M}_\odot$). More generally, the empirical evidence for large tracer-dependent offsets in disk kinematics for high redshift galaxies remains inconclusive. The most direct test is to compare multiple tracers within the same galaxies. When this has been done, no clear systematic offset in $V_{\rm rot}/\sigma_0$ between different gas phases has been found \citep{Lee+25,wisnioski+25}.

It remains unclear which ISM phase predominantly gives rise to the observed \cii emission, and whether the relative contributions from different ISM phases vary with underlying galaxy properties \citep{wisnioski+25}. Another possibility is that the \cii-based velocity dispersions are overestimated due to the modest spatial resolution of the interferometric data ($\sim 2-4$\,kpc; see \citealt{Lee+25} for more details).

Overall, the galaxy disks identified here display ordered rotation with moderate values of $V_{2.2}/\sigma_0 \approx 1$, indicative of dynamically warm systems. We note that these estimates are consistent with the JWST/slitless (grism) measurements made for a larger population of $\sim100$ galaxies at $z = 4-6$ \citep[][$\sigma_0 \approx 100$\,km\,s$^{-1}$ and $V_{\rm rot}/\sigma_0 \approx 1.0$]{Danhaive+25}. Our measurements suggest that these early disks are not yet thin, cold disks like those we commonly observe at low redshift \citep{Epinat+10}. Instead, they appear thick and turbulent, likely sustained by rapid gas accretion and intense feedback events \citep{Krumholz+18}. The fact that we detect a significant disk fraction at $z = 4-6$, and that their kinematics can be understood under the framework of gravitationally marginally stable turbulent disks, provides clues that disk settling is already underway. Previous works have already suggested that massive galaxies ($M_\star \gtrsim 10^{10}\,{\rm M}_\odot$) at $z \approx 4-5$ can already be settled on rotating disks \citep{Neeleman+20,Lelli+21}. In contrast, the disk candidates identified in this work are less massive, with $M_\star \gtrsim 10^{9.5}\, {\rm M}_\odot$. The broad diversity of kinematics and morphologies seen in the full ALPINE-CRISTAL-JWST sample suggests that disk settling is an emerging process at this cosmic epoch and over a wider range of stellar mass than previously thought.

\section{Conclusions}
\label{sec:conclusions}

We have presented a first analysis of the stellar mass and baryonic Tully-Fisher relations at $z=4-6$ using JWST/NIRSpec IFU observations from the ALPINE-CRISTAL-JWST survey, complemented with JWST/NIRCam imaging. Our main results can be summarized as follows:

\begin{enumerate}

\item We identify five out of 18 galaxies as isolated disk-like systems, corresponding to a sample disk fraction of $f_{\rm disk}=0.28\pm0.11$ at $z=4-6$. In addition, four systems are classified as galaxy pairs, while the remaining nine show perturbed kinematics.

\item The five disk-like galaxies show clear velocity gradients and rotation curves that extend up to or beyond $\sim2.2R_{\rm d}$, allowing estimates of their kinematic properties. The sample spans { intrinsic $V_{2.2}=91-127\,{\rm km\,s^{-1}}$}, { $\sigma_0=66-129\,{\rm km\,s^{-1}}$}, and { $V_{2.2}/\sigma_0=0.9-1.9$}, suggesting that random motions remain important in these systems. Our measured intrinsic velocity dispersions and $V_{2.2}/\sigma_0$ ratios are consistent with the expectations for marginally stable disks given the average gas fraction of the systems at this cosmic epoch.

\item We find that the disk galaxies in our sample are offset with respect to the low-$z$ stellar-mass and baryonic Tully-Fisher relations. In the sTFR, all five disks lie below the local relation and the one at $z\approx2.3$, suggesting zero-point offsets of { $-0.96\pm0.18$\,dex} and { $-0.50\pm0.18$\,dex}, respectively. In the bTFR, all five disks lie below the local relation, while three are broadly consistent with the $z\approx2.3$ bTFR. We estimate zero-point offsets of { $-0.57\pm0.20$\,dex} relative to the local bTFR and { $-0.30\pm0.20$\,dex} relative to the $z\approx2.3$ relation.

\item We extend the empirical model of \citet{Ubler+17} up to $z=5$, finding good agreement with our best-fit zero-points of both the sTFR and bTFR. In this framework, the redshift evolution of both TFR zero-points is driven by the cosmic evolution of dark matter halo properties, modulated by a reduced contribution of dark matter to the dynamical mass on the galaxy scale.

\end{enumerate}

Overall, our results suggest that rotating disk-like systems were already in place within the first $\sim1$\,Gyr of cosmic time, although their kinematics show high degree of turbulence. While our conclusions are limited by small number statistics, this work provides one of the first spatially resolved views of the Tully-Fisher relation for main-sequence galaxies at $z=4-6$. Larger JWST/NIRSpec IFU samples, together with deeper ALMA observations for the cold gas and spatially resolved dynamical modeling, will be key for determining whether the offsets reported here are representative of the bulk galaxy population at these redshifts.

\begin{acknowledgements}
We thank the anonymous referee for their careful reading and constructive comments, which greatly improved the quality and clarity of this manuscript.

We warmly thank Annagrazia Puglisi and Andr\'es Escala for their comments, which helped to improve this study. D.G.E. acknowledges support from ANID BECAS/DOCTORADO NACIONAL/2024-21251071. D.G.E., J.M., E.I., M.A., R.H.C and V.V. gratefully acknowledge support from ANID MILENIO NCN2024$\_$112, D.G.E. and E.I. acknowledge support from ANID FONDECYT Regular 1221846.

M.A. is supported by FONDECYT grant number 1252054.

M.A. acknowledges support from ANID+Vinculaci\'on Internacional + FOVI250261.

{ M.A.} M.B., RJA, and V.V. acknowledge support from the ANID BASAL project FB210003. This work was supported by the French government through the France 2030 investment plan managed by the National Research Agency (ANR), as part of the Initiative of Excellence of Université Côte d’Azur under reference number ANR-15-IDEX-01. This research was funded, in whole or in part, by the French National Research Agency (ANR), grant ANR-25-CE31-0192 (project CIGALE2D)

RJA was supported by FONDECYT grant number 1231718.

EdC acknowledges support from the Australian Research Council through project DP240100589.

M. Palla acknowledges support from HORIZON-INFRA-2024-DEV-01-01 – Research Infrastructure Concept Development, through the project WST: The Wide-Field Spectroscopic Telescope (Grant No. 101183153).

M. Relano acknowledges support from project PID2023-150178NB-I00 financed by MCIU/AEI/10.13039/501100011033, and by FEDER, UE.

HÜ thanks the Max Planck Society for support through the Lise Meitner Excellence Program. H\"U acknowledges funding by the European Union (ERC APEX, 101164796). Views and opinions expressed are however those of the authors only and do not necessarily reflect those of the European Union or the European Research Council Executive Agency. Neither the European Union nor the granting authority can be held responsible for them.

\end{acknowledgements}

%
%

\bibliographystyle{aa}   
\bibliography{references} 

\begin{appendix}

\section{Global properties of the ALPINE-CRISTAL-JWST sample}
\label{app:globalprop}

In Table~\ref{tab1}, we present a compilation of different properties for the full ALPINE-CRISTAL-JWST sample.

\begin{table*}
    \def\arraystretch{1.3}
    \setlength\tabcolsep{3pt}
    \caption{Global properties of the ALPINE-CRISTAL-JWST sample.}
    \vspace{-0.2mm}
    \label{tab1}
    \small
    \centering
    \renewcommand{\arraystretch}{1.3}
    \begin{tabular}{lcccccccc}
        \hline
        ID & CRISTAL ID & R.A. & Decl. & $z_{H\alpha}$ & $\log M_\star$ & $\log \mathrm{SFR}_{\mathrm{H}\alpha}$ & $\log L_{\rm [CII]}$ & $\log M_{\rm gas}$ \\
        & & (J2000.0) & (J2000.0) & & $({\rm M}_\odot)$ & $({\rm M}_\odot\,\mathrm{yr}^{-1})$ & $(L_\odot)$ & $({\rm M}_\odot)$ \\
        (1) & (2) & (3) & (4) & (5) & (6) & (7) & (8) & (9) \\
        \hline
        DC--417567 & C--10 & 10:02:04.08 & +01:55:44.0 & 5.6700 & $9.99 \pm 0.31$ & $1.86 \pm 0.02$ & $8.50 \pm 0.08$ & $9.98 \pm 0.32$ \\
        DC--494763 & C--19 & 10:00:05.10 & +02:03:12.2 & 5.2337 & $9.51 \pm 0.36$ & $1.76 \pm 0.03$ & $8.69 \pm 0.05$ & $10.17 \pm 0.31$ \\
        DC--519281 & C--09 & 09:59:00.89 & +02:05:27.6 & 5.5759 & $9.84 \pm 0.39$ & $2.12 \pm 0.05$ & $8.73 \pm 0.09$ & $10.21 \pm 0.32$ \\
        DC--536534 & C--03 & 09:59:53.24 & +02:07:05.4 & 5.6886 & $10.40 \pm 0.29$ & $2.61 \pm 0.04$ & $8.90 \pm 0.09$ & $10.39 \pm 0.32$ \\
        DC--630594 & C--11 & 10:00:32.60 & +02:15:28.4 & 4.4403 & $9.68 \pm 0.32$ & $1.83 \pm 0.08$ & $8.79 \pm 0.04$ & $10.27 \pm 0.31$ \\
        DC--683613 & C--05 & 10:00:09.42 & +02:20:13.8 & 5.5420 & $10.16 \pm 0.35$ & $1.60 \pm 0.05$ & $8.90 \pm 0.04$ & $10.39 \pm 0.31$ \\
        DC--709575 & C--14 & 09:59:47.06 & +02:22:32.8 & 4.4121 & $9.53 \pm 0.38$ & $1.36 \pm 0.06$ & $8.48 \pm 0.09$ & $9.96 \pm 0.32$ \\
        DC--742174 & C--17 & 10:00:39.12 & +02:25:32.2 & 5.6360 & $9.51 \pm 0.40$ & $1.18 \pm 0.06$ & $8.17 \pm 0.10$ & $9.64 \pm 0.32$ \\
        DC--848185 & C--02 & 10:00:21.51 & +02:35:10.9 & 5.2931 & $10.30 \pm 0.28$ & $2.43 \pm 0.03$ & $9.20 \pm 0.03$ & $10.69 \pm 0.31$ \\
        DC--873321 & C--07 & 10:00:04.06 & +02:37:35.8 & 5.1542 & $10.00 \pm 0.33$ & $2.05 \pm 0.03$ & $8.98 \pm 0.05$ & $10.47 \pm 0.31$ \\
        DC--873756 & C--24 & 10:00:02.72 & +02:37:40.1 & 4.5457 & $10.53 \pm 0.08$ & $0.30 \pm 0.50$ & $9.56 \pm 0.02$ & $11.06 \pm 0.31$ \\
        DC--842313 & C--01 & 10:00:54.53 & +02:34:35.0 & 4.5537 & $10.65 \pm 0.50$ & $2.50 \pm 0.14$ & $8.66 \pm 0.08$ & $10.14 \pm 0.32$ \\
        VC--5100541407 & C--06 & 10:01:00.92 & +01:48:33.7 & 4.5630 & $10.09 \pm 0.30$ & $0.70 \pm 0.49$ & $9.07 \pm 0.04$ & $10.56 \pm 0.31$ \\
        VC--5100822662 & C--04 & 09:58:57.91 & +02:04:51.4 & 4.5205 & $10.15 \pm 0.29$ & $1.83 \pm 0.08$ & $8.90 \pm 0.04$ & $10.39 \pm 0.31$ \\
        VC--5100994794 & C--13 & 10:00:41.14 & +02:17:14.1 & 4.5802 & $9.65 \pm 0.33$ & $1.89 \pm 0.13$ & $8.75 \pm 0.04$ & $10.23 \pm 0.31$ \\
        VC--5101218326 & C--25 & 10:01:12.50 & +02:18:52.6 & 4.5739 & $10.90 \pm 0.32$ & $1.04 \pm 0.49$ & $9.26 \pm 0.02$ & $10.75 \pm 0.31$ \\
        VC--5101244936 & C--15 & 10:00:47.65 & +02:18:02.1 & 4.5803 & $9.69 \pm 0.33$ & $1.88 \pm 0.09$ & $8.70 \pm 0.09$ & $10.18 \pm 0.32$ \\
        VC--5110377875$^{(*)}$ & \ldots & 10:01:32.34 & +02:24:30.3 & 4.5505 & $10.17 \pm 0.17$ & $1.96 \pm 0.17$ & $9.23 \pm 0.03$ & $10.72 \pm 0.31$ \\
        \hline
    \end{tabular}
    \renewcommand{\arraystretch}{1.0}
    \tablefoot{
    (1) ALPINE source name \citep{Lefevre+20}. (2) CRISTAL survey name \citep{Herrera-Camus+25}. (3) Right ascension. (4) Declination. (5) Redshift. (6) Stellar mass taken from \citet{Faisst+26}. (7) Star formation rate estimated from H$\alpha$ emission-line luminosity, corrected for dust attenuation \citep{Faisst+26}. (8) [C\,{\sc ii}] emission-line luminosity \citep{Bethermin+20}. (9) Cold-gas mass inferred from [C\,{\sc ii}] luminosity using the prescription of \citet{Zanella+18}. ($^*$) This system was not covered by the CRISTAL survey.
    }
\end{table*}

\section{Comparison of datasets for CRISTAL-09}
\label{app:1}

\begin{figure}
    \centering
    \includegraphics[width=0.95\linewidth]{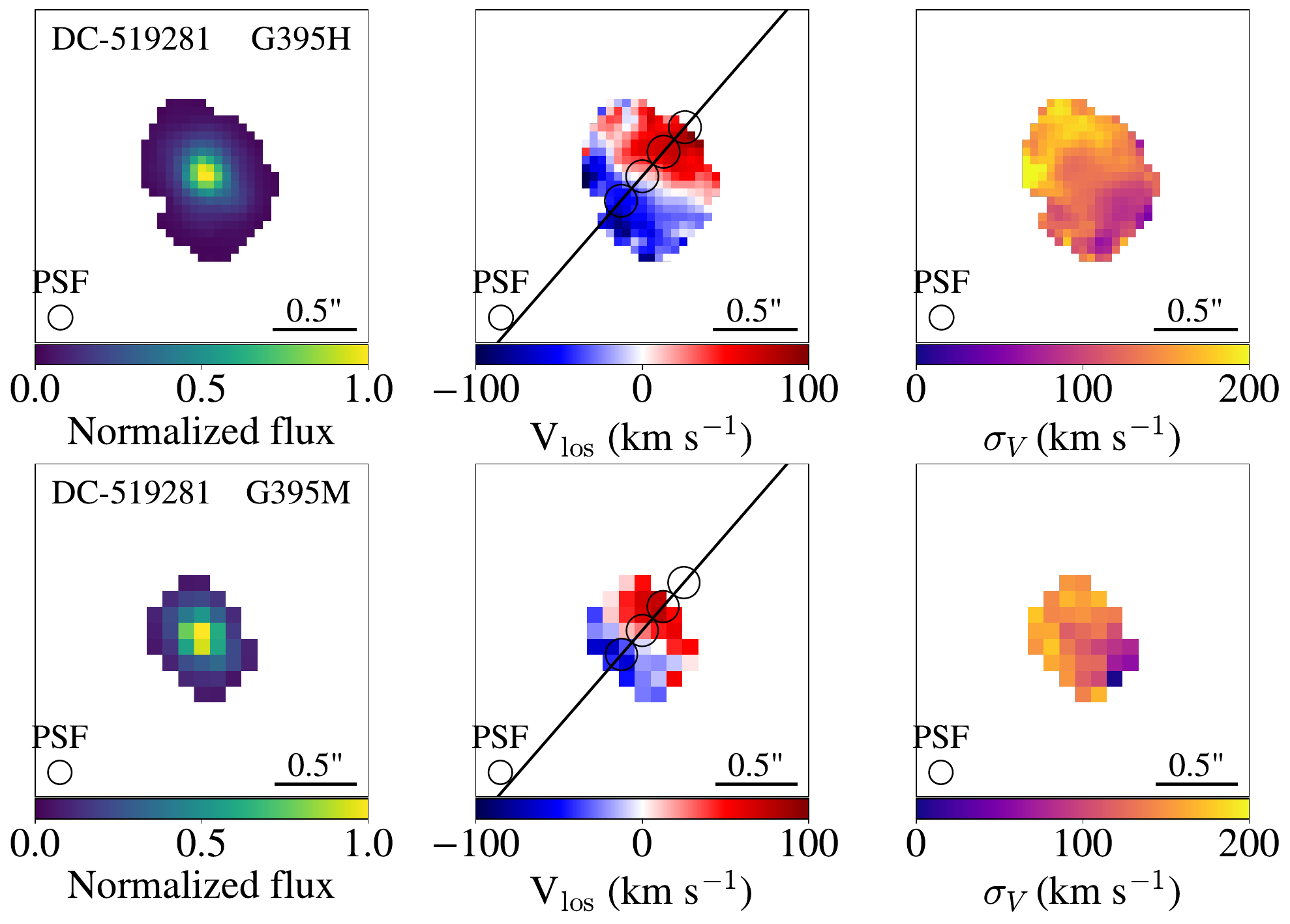}
    \caption{Comparison of the H$\alpha$ moment maps for DC--519281/C--09, obtained from the G395H (\textit{top}) and G395M (\textit{bottom}) NIRSpec-IFU grating data. The maps are color coded following Figure~\ref{fig:summary_5x5}.}
    \label{fig:comparison}
\end{figure}

\begin{figure}
    \centering
    \includegraphics[width=1\linewidth]{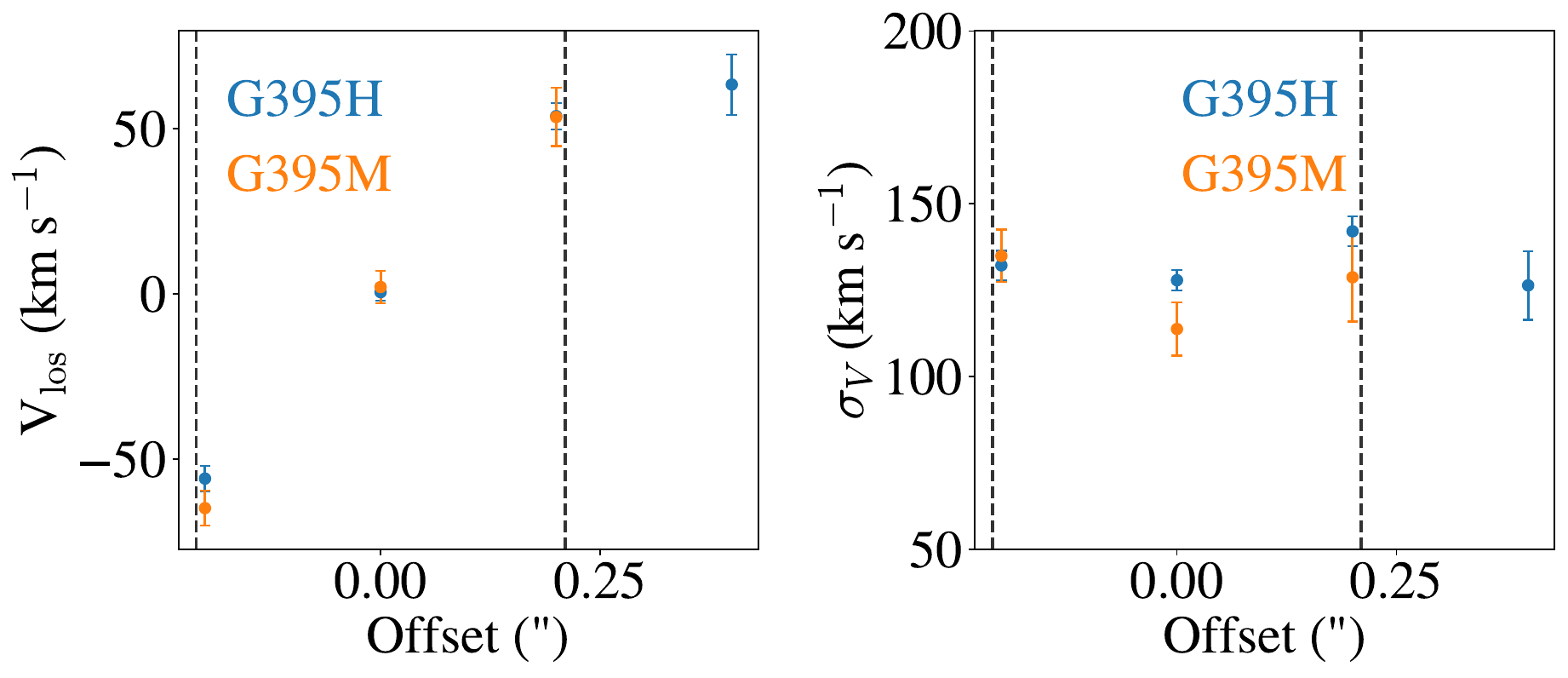}
    \caption{Comparison of the rotation curve (\textit{left}) and velocity dispersion radial profile (\textit{right}) for DC--519281/C--09, obtained from the G395H and G395M NIRSpec-IFU grating data. In both panels, the dashed line marks the extraction radius $2.2\,R_{\rm d}$.} 
    \label{fig:kin_comparison}
\end{figure}

The available ORCHIDS observations for DC--519281/C--09 allow us to assess whether the ALPINE-CRISTAL-JWST data are sufficient for recovering the intrinsic velocity dispersion values for this source. In Figure~\ref{fig:comparison} we present a comparison between these two datasets. { The ORCHIDS observations use G395H/F290LP ($R\approx2700$) with a spatial sampling of $0.05''$\,pixel$^{-1}$, while the ALPINE-CRISTAL-JWST observations use G395M ($R\approx1000$) with $0.10''$\,pixel$^{-1}$ sampling.} Overall, we find { broad} agreement. The moment maps show consistent features among both datasets, despite the enhanced data quality delivered by the observing setup of ORCHIDS. { For a consistent comparison, the kinematic profiles are extracted using circular apertures of $0.2''$ in diameter and separated by $0.2''$ along the same kinematic major axis. The aperture positions are defined from the G395H data with S/N$>5$ and the same positions are used for G395M. One of the four G395H apertures has S/N$<5$ in G395M and is therefore not included in the G395M profile.} This agreement can be quantitatively assessed in Figure~\ref{fig:kin_comparison}, { where the rotation velocities agree within $1\sigma$ at two of the three common aperture positions, and the velocity dispersions independently agree within $1\sigma$ at two of the three positions.} The $V_{\rm los}$ extracted at $2.2\,R_{\rm d}$ is { $55 \pm 3$\,km\,s$^{-1}$ for the G395H grating and $55 \pm 5$\,km\,s$^{-1}$ for the G395M grating. The velocity dispersion at $2.2R_d$ is $128 \pm 3$\,km\,s$^{-1}$ for G395H and $113 \pm 7$\,km\,s$^{-1}$ for G395M.} The uncertainties tend to be smaller in the high resolution data.

\section{Exponential profiles for the full sample}
\label{app:exponential_models}

{ Figure~\ref{fig:appendix_pysersic1} shows the F444W images, the adopted exponential ($n=1$) {\sc PySersic} models, and the corresponding residuals for the full sample.}

\begin{figure*}
    \centering
    \includegraphics[width=1\linewidth]{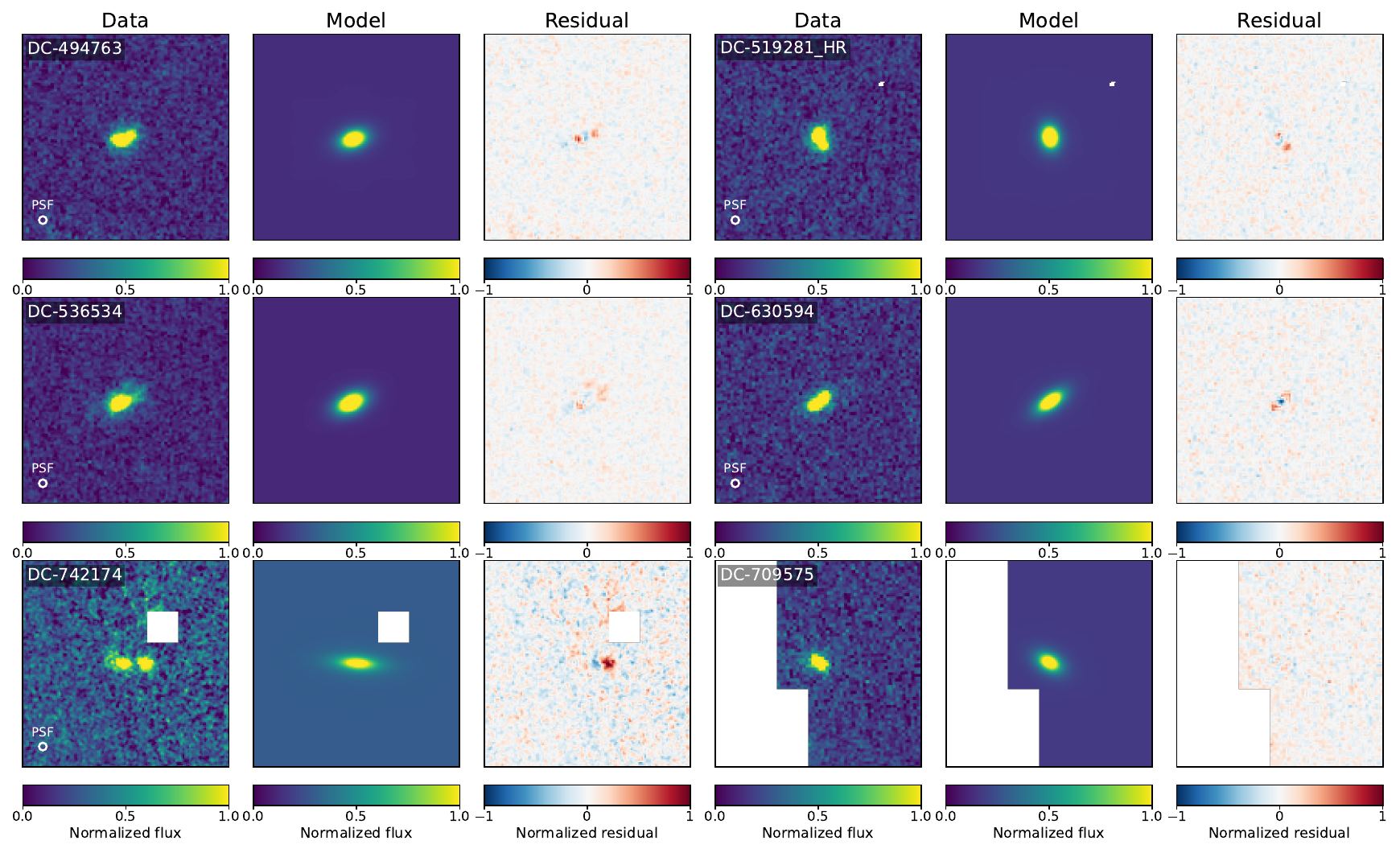}
    \caption{F444W imaging, best-fit exponential models, and residuals for the galaxies in our sample. The white pixels represent masked regions due to instrumental features. The residuals are the data–model difference normalized by the image flux scale.} 
    \label{fig:appendix_pysersic1}
\end{figure*}

\begin{figure*}
    \ContinuedFloat
    \centering
    \includegraphics[width=1\linewidth]{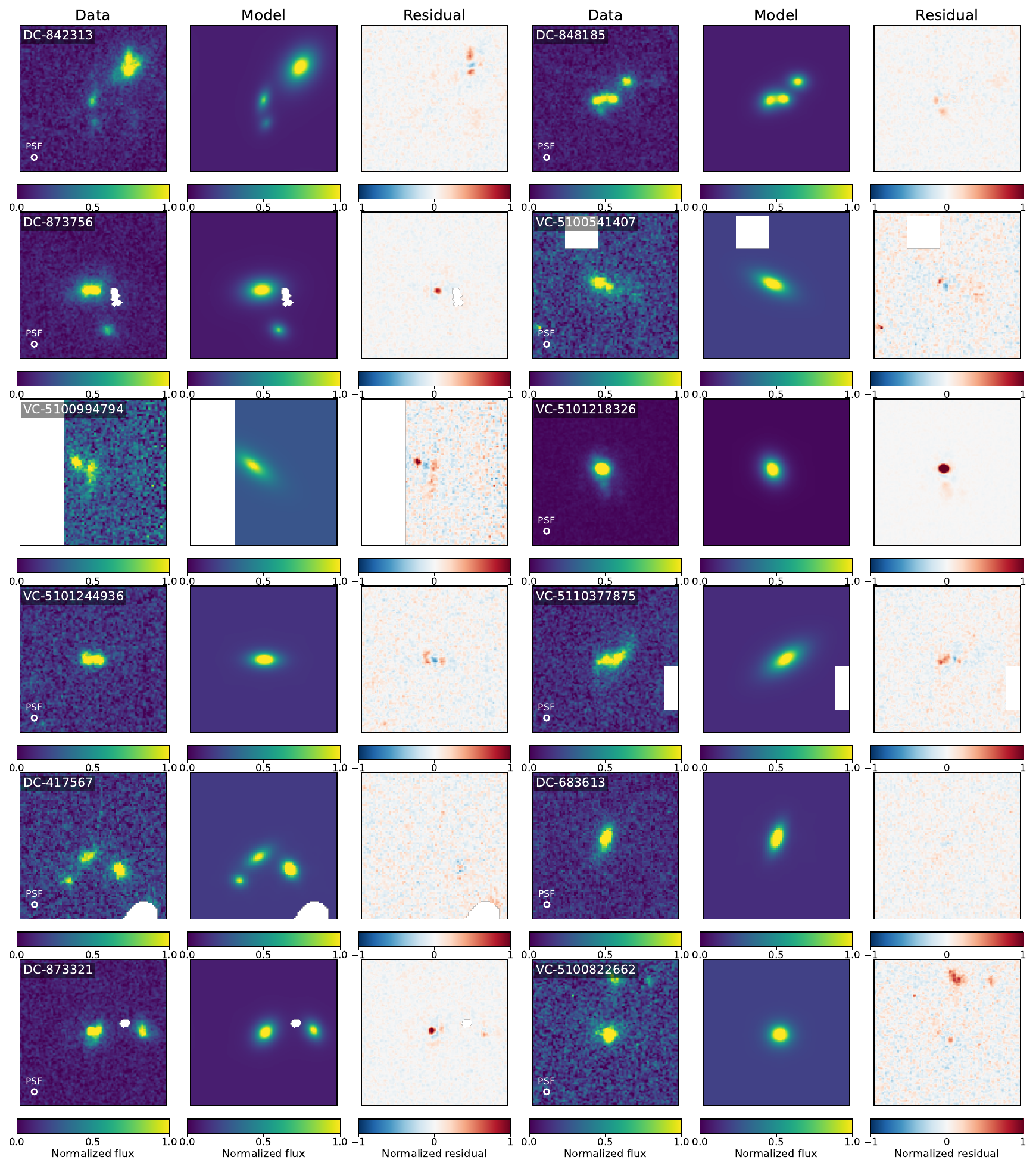}
    \caption{Continued}
    \label{fig:appendix_pysersic2}
\end{figure*}

\section{Non-disks and galaxy pairs}
\label{app:allsample}

For completeness, Figure~\ref{nondisks_image1} shows the JWST false-colour images and H$\alpha$ moment maps for the non-disk systems and galaxy pairs.

\begin{figure*}
    \centering
    \includegraphics[width=0.75\linewidth]{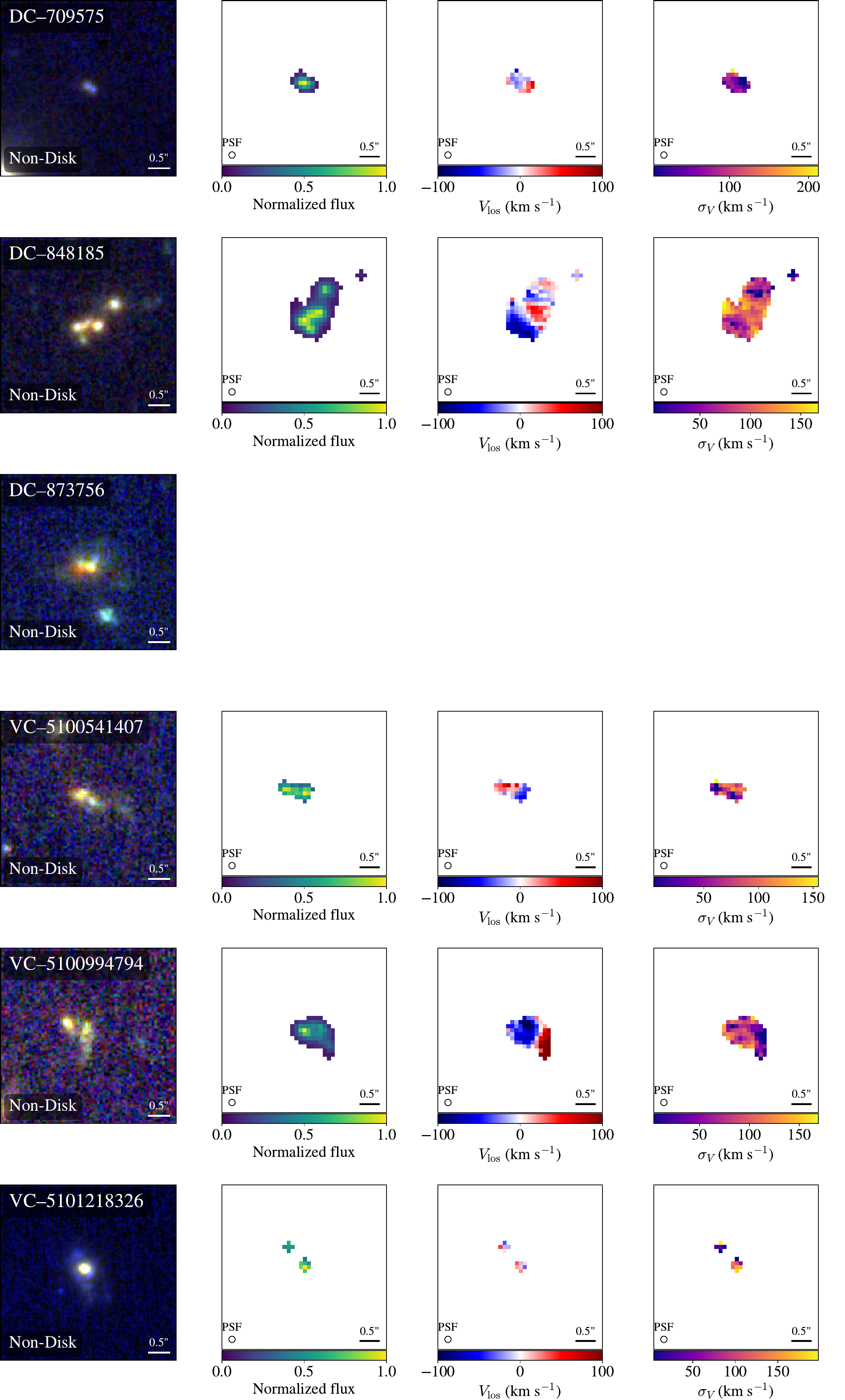}
    \caption{Summary of the galaxies classified as non-disks or pairs in this work, excluding the disk sample shown in Fig.~\ref{fig:summary_5x5}. Each row shows one galaxy, with the classification indicated in the lower-left corner of the JWST/NIRCam panel. From left to right, the panels show the JWST/NIRCam RGB image, the normalized H$\alpha$ line-intensity map, velocity field, and velocity dispersion map. Objects without available JWST imaging or without a significant ($5\sigma$) H$\alpha$ detection are indicated in the corresponding panel.}
    \label{nondisks_image1}
\end{figure*}

\begin{figure*}
    \ContinuedFloat
    \centering
    \includegraphics[width=0.75\linewidth]{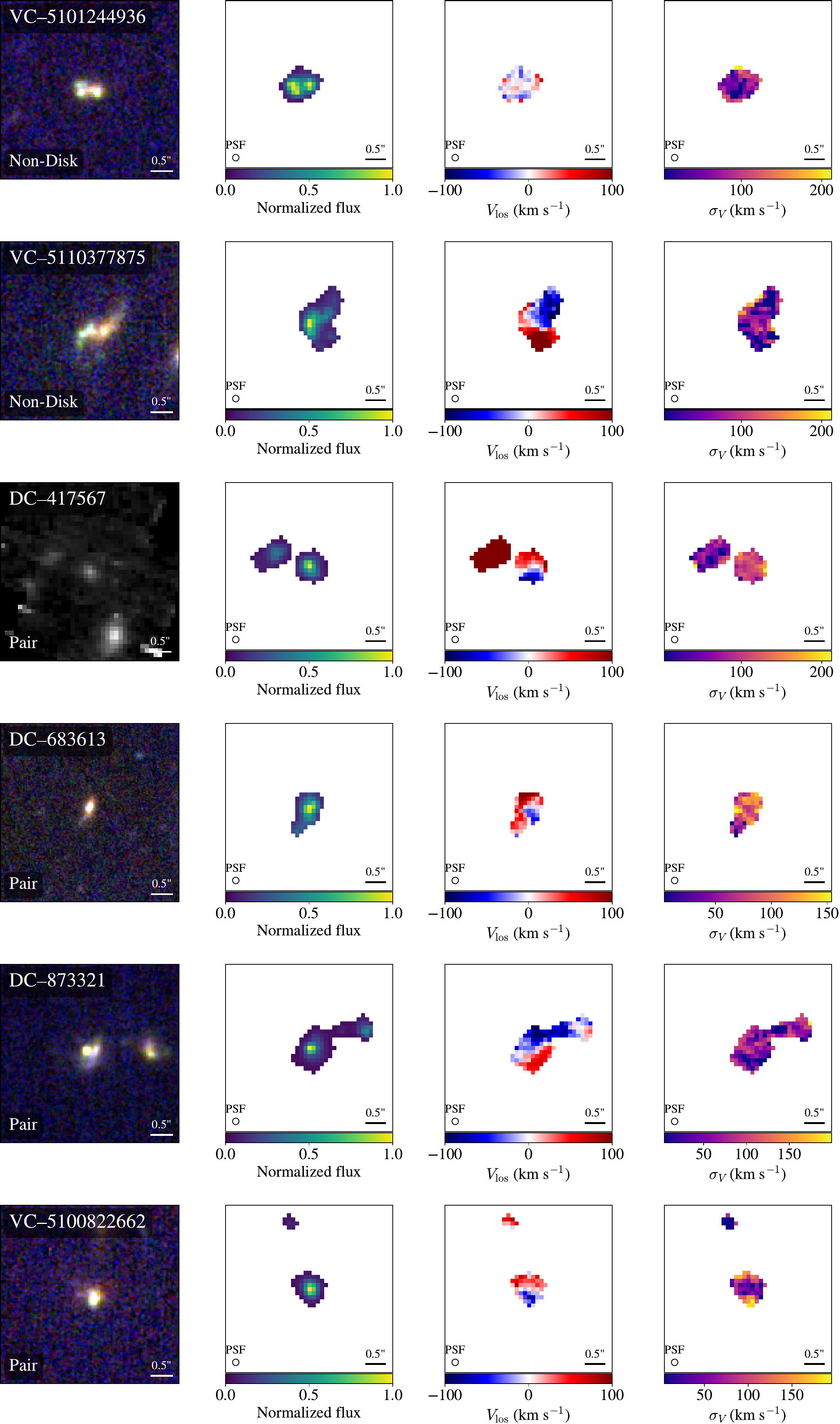}
    \caption{continued. Due to the lack of F444W image for DC--417567, we show a pseudo image made from the NIRSpec/IFU data.}
    \label{fig:nondisks_image_cont}
\end{figure*}

\section{Recovery test for $\sigma_V$}
\label{app:vsig_sim}

Some of the velocity dispersion values for our sources are lower than the LSF width of the NIRSpec-IFU G395M grating ($\sigma_{\rm LSF} \approx 127~\kms$). This is because many observed emission lines are detected over several consecutive spectral channels due to their high S/N. Nevertheless, it is important to determine the minimum S/N at which we can reliably recover $\sigma_V$ given the limited spectral resolution ($R \approx 1000$).

\begin{figure}
    \centering
    \includegraphics[width=0.9\linewidth]{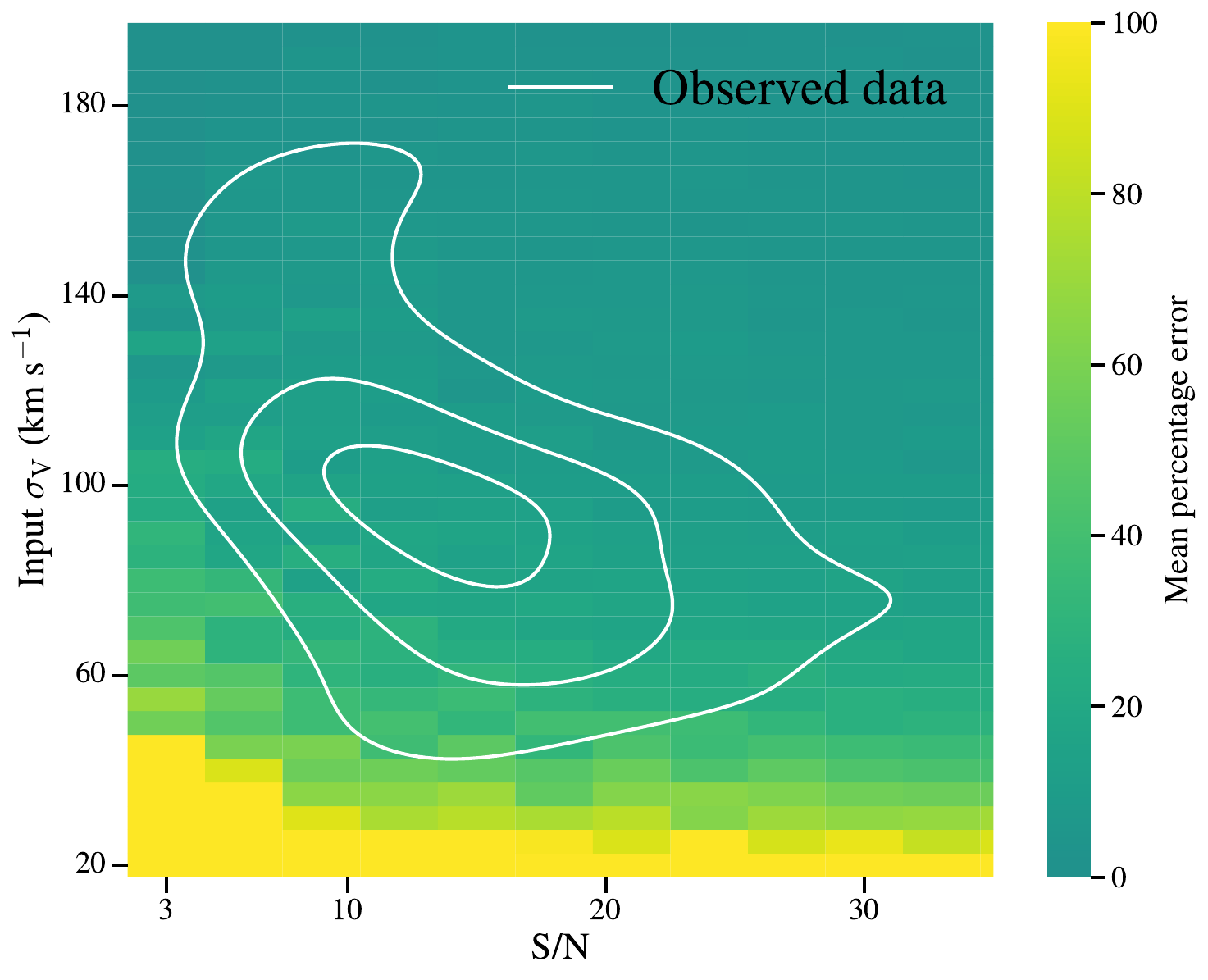}
    \caption{Mean percentage error of the recovered velocity dispersion as a function of the intrinsic (input) value and S/N, derived from our simulations. The white contours encompass the 50, 70 and 80$\%$ of the pixels in the sample.}
    \label{fig:grid}
\end{figure}

To address this, we performed a set of simulations adapted to the properties of the data for each galaxy. First, we generated synthetic H$\alpha$ emission lines, assuming Gaussian profiles and adopting different intrinsic velocity dispersion values. Then, the simulated emission lines were shifted to the redshift of each galaxy, convolved with the instrumental LSF \citep{Shajib+25}, and resampled to match the spectral step of the JWST data cube. To reproduce realistic observational conditions, we added noise extracted directly from emission-line-free spaxels in the data cubes. We produced many mock emission lines with varying intrinsic velocity dispersions and S/N ratios, considering multiple noise extraction realizations. Finally, these mock spectra were analyzed using the same procedure applied to the observed emission lines (Section~\ref{sec:methods}). Figure~\ref{fig:grid} compares the intrinsic velocity dispersion with the recovered values. The G395M grating data quality is insufficient to recover emission lines with $\sigma_V \lesssim 40$\,km\,s$^{-1}$, regardless of the observed S/N. In the regime most relevant for our data (S/N$\,\approx 5$--20 and intrinsic velocity dispersions $\sim 70$--$100\,$km\,s$^{-1}$), we find that the velocity dispersion can be recovered with an accuracy $\sim 20-40\%$. This result indicates that we can reliably recover $\sigma_V$ for our systems and is consistent with the comparison between different JWST NIRSpec-IFU observing setups for DC--519281.

\begin{figure}
    \centering
    \includegraphics[width=0.85\linewidth]{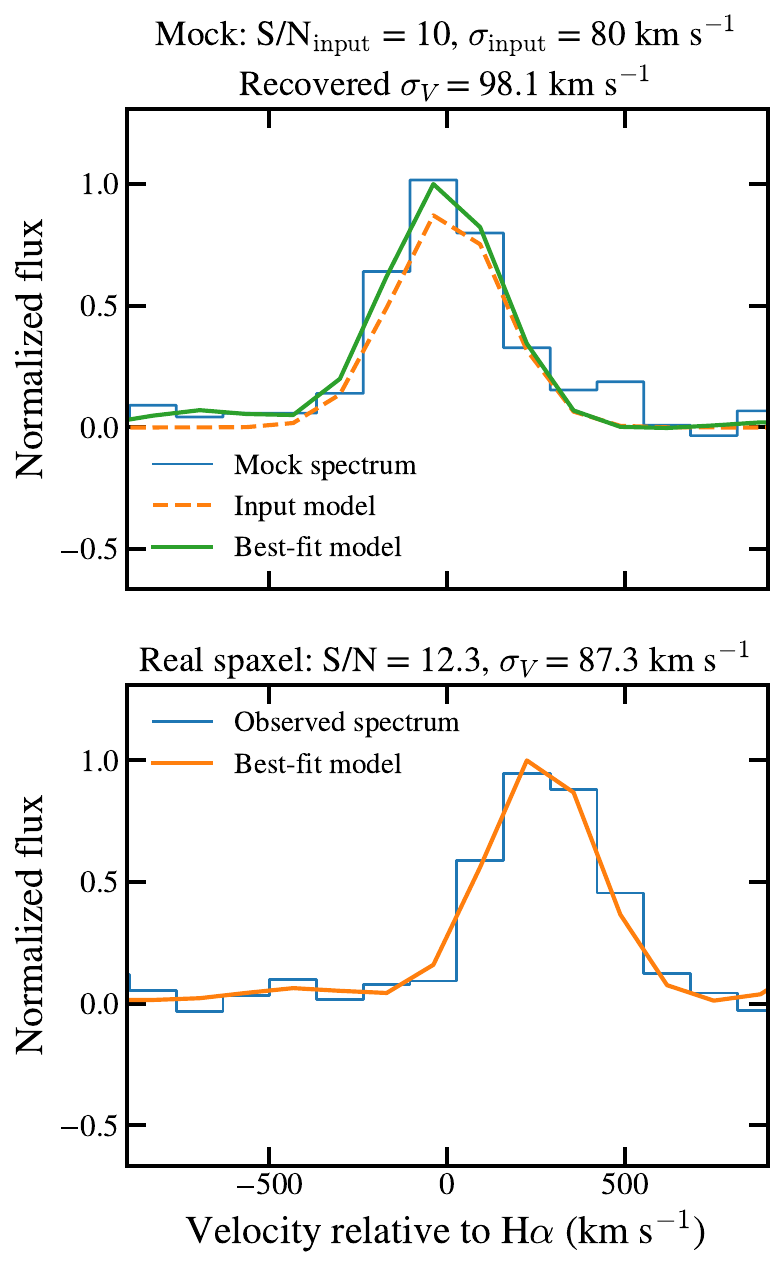}
    \caption{Top: simulated H$\alpha$ spectrum with S/N = 10 and an intrinsic velocity dispersion of $\sigma_V = 80$ km s$^{-1}$, together with the input and best-fit models. Bottom: observed spectrum from a spaxel with similar S/N and velocity dispersion, together with its best-fit model.} 
    \label{fig:mock_spectra}
\end{figure}

\section{Asymmetric-drift correction}
\label{app:adc}

Given the comparable values of $V_{2.2}$ and $\sigma_0$ in our sample, asymmetric-drift corrections are significant: the circular velocities are obtained by applying factors of $\sim2{-}3$ to the rotational velocities. However, the magnitude of the asymmetric-drift corrections depends on estimates for the pressure gradient within galactic disks, which are based on assumptions about the gas surface density, velocity dispersion, and disk vertical structure (see appendix in \citealt{Bouche+22}). These estimates are difficult to constrain at high redshift. 

Here, we test how different asymmetric-drift prescriptions affect the inferred stellar Tully-Fisher offset. We start by re-writing Equation~\ref{eq:ADC} as

\begin{equation}
V_{\rm circ}^2(R) = V_{\rm rot}^2(R) + \alpha \sigma_0^2\left(\frac{R}{R_{\rm d}} \right) ,
\end{equation}

\noindent where $\alpha$ is the factor that varies depending on the adopted prescription. A galaxy model with a constant scale height gives $\alpha=1$ \citep{Bouche+22}. For thick-disk turbulent galaxies with mass distributions following flattened S\'ersic profiles, we have $\alpha \approx 1.3$ for a self-consistent asymmetric-drift correction \citep{Price+22}. Finally, the \citet{Burkert+10} prescription adopted in our main analysis gives $\alpha=2$, which assumes an exponential gas surface density profile and a constant, isotropic velocity dispersion.

\begin{figure}
    \centering
    \includegraphics[width=0.85\linewidth]{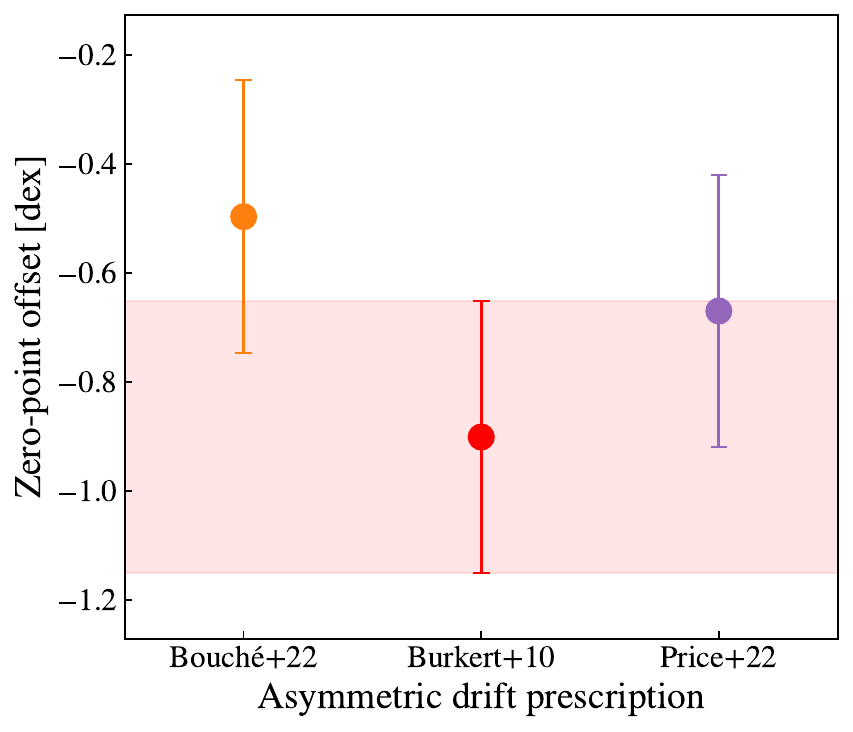}
    \caption{Stellar Tully-Fisher offset relative to the local relation of \citet{Reyes+11} for different pressure-support prescriptions. Each point shows the zero-point offset obtained from a fixed-slope fit to our five disk galaxies. We show the constant scale-height case \citep{Bouche+22}, the \citet{Price+22} correction, and \citet{Burkert+10}'s correction.}
    \label{fig:appendix_ADC}
\end{figure}

Fig.~\ref{fig:appendix_ADC} shows the stellar Tully-Fisher offset relative to \citet{Reyes+11} for these three prescriptions. Lower values of $\alpha$ produce less negative offsets, as expected. However, all prescriptions agree within the uncertainties. We adopt the \citet{Burkert+10} correction because it is consistent with our use of a single characteristic dispersion, $\sigma_0$, and allows a direct comparison with previous high-redshift Tully-Fisher studies.

\section{Cosmic evolution of $f_{\rm DM}(< R_e)$}  
\label{app:fdm}

\begin{figure*}
    \centering
    \includegraphics[width=0.8\linewidth]{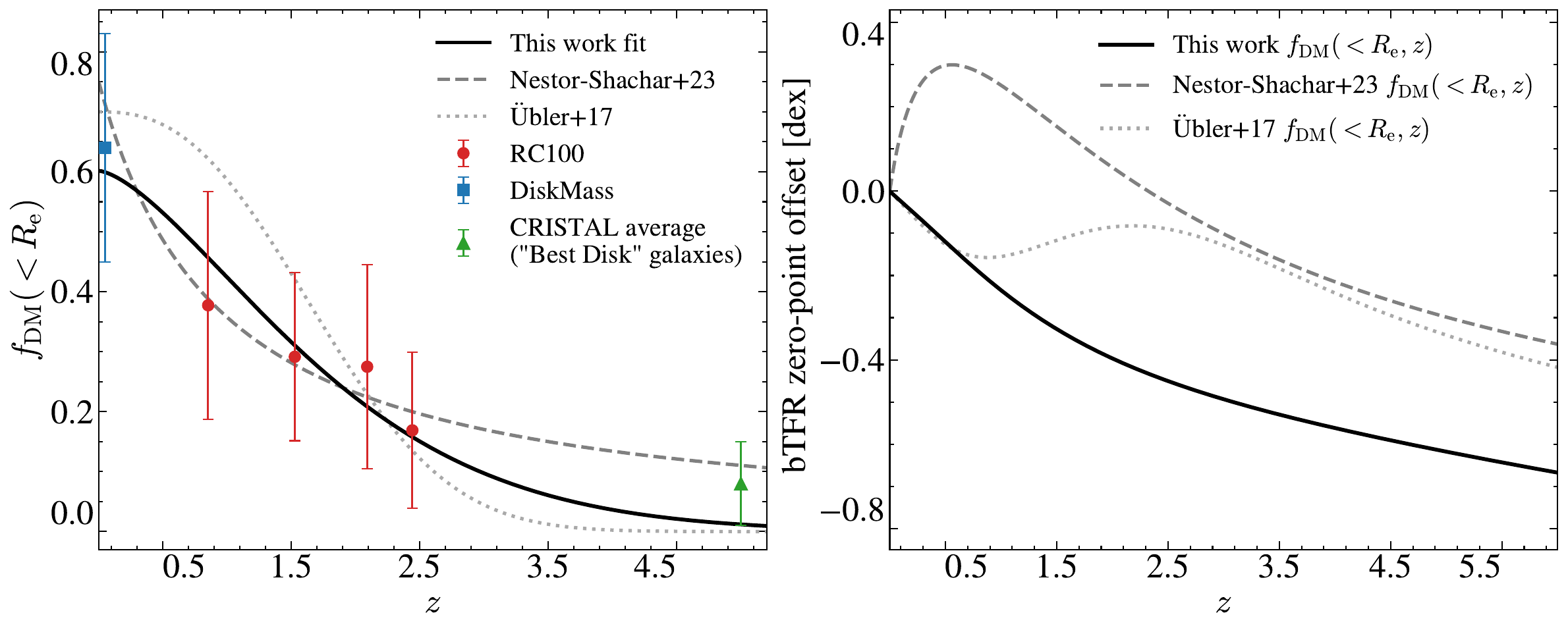}
    \caption{\textit{Left}: redshift evolution of the dark matter fraction within one effective radius, $f_{\rm DM}(<R_{\rm e},z)$. \textit{Right}: evolution of the baryonic TFR zero point predicted by the model of \citet{Ubler+17} when adopting different parameterizations for $f_{\rm DM}(<R_{\rm e},z)$.}
    \label{fig:fdm_appendix}
\end{figure*}

The redshift evolution of $f_{\rm DM}(<R_{\rm e},z)$ is the least constrained variable in the model of \citet{Ubler+17}. Here, we derive a functional form for $f_{\rm DM}(<R_{\rm e},z)$ by fitting measurements from the literature. At $z=0$, we consider $f_{\rm DM} = 0.64 \pm 0.18$ from the DiskMass Survey \citep{Martinsson+13}. For the $z = 0.7 - 2.5$ range, we adopt the measurements for the RC100 sample published in \citet{Nestor-Shachar+23}. It is worth noting that the stellar-mass ranges of those comparison samples are slightly higher than the range probed in this work, $10^{10-11}\,{\rm M}_\odot$ versus $10^{9.5-10.5}\,{\rm M}_\odot$. Nevertheless, we adopt these samples because they currently represent the lowest-mass systems available in the literature with reliable $f_{\rm DM}$ determinations at $z = 1-3$. Additionally, we take the average $f_{\rm DM} = 0.08 \pm 0.07$ value derived for our disk galaxies in \citep{Lee+25}. During the fitting, we adopt the following parameterization:

\begin{equation}
    f_{\rm DM}(<R_{\rm e},z) = A \exp\left(-B \, z^{C}\right),
    \label{eq:fdm_param}
\end{equation}
where $A$, $B$, and $C$ are free parameters. We find
\begin{equation}
    A = 0.58 \pm 0.08; \qquad
    B = 0.32 \pm 0.19; \qquad
    C = 1.50 \pm 0.54.
\end{equation}

We show the best-fit results in Figure~\ref{fig:fdm_appendix}. By comparing with the power-law parameterization of \citet{Nestor-Shachar+23} for the RC100 sample, we find that our best-fit predicts lower $f_{\rm DM}(<R_{\rm e},z)$ values at $z\gtrsim 2.5$. At low-$z$, our best-fit also predicts lower dark matter fractions, but the $f_{\rm DM}(<R_{\rm e},z)$ trend does not peak at high values as in the parameterizations of \citet{Nestor-Shachar+23}. The parameterization of \citet{Ubler+17} considers higher $f_{\rm DM}(<R_{\rm e})$ values for galaxies at low-$z$, but it broadly agrees with ours at $z \gtrsim 2.3$

In practice, the $f_{\rm DM}(<R_{\rm e})$ data show no preference for any model due to the significant uncertainties. However, compared with the parameterization of \citet{Nestor-Shachar+23} our best-fit $f_{\rm DM}(<R_{\rm e},z)$ induces a substantially different prediction for the bTFR (and sTFR) zero-point evolution with redshift in the model of \citet{Ubler+17}. The parameterization of \citet{Nestor-Shachar+23} implies positive offsets for the bTFR at low-$z$, while our best-fit does not. This is because the $f_{\rm DM}(<R_{\rm e})$ parameterization of \citet{Nestor-Shachar+23} accounts for a significant reduction in the dark matter content within galaxies at $z \lesssim 1$, which counterbalances and further dominates the effect induced by the $H(z)^{-1}$ term in the model of \citet[][see Equation~\ref{eq:toy_ubler}]{Ubler+17}. This effect can also be seen when adopting the $f_{\rm DM}(<R_{\rm e})$ parameterization of \citet{Ubler+17}. The positive zero-point evolution of the bTFR (and sTFR) described by assuming the $f_{\rm DM}(<R_{\rm e})$ parameterization \citet{Nestor-Shachar+23} is not consistent with the observational data at $z \lesssim 1$, which is why we prefer our analytical prescription for $f_{\rm DM}(<R_{\rm e},z)$.

\end{appendix}

\end{document}